\documentclass[pre,twocolumn,twoside,showpacs,superscriptaddress,tightenlines]{revtex4-1}
\usepackage{graphicx,amssymb,amsmath,epsf,bm,amsfonts}
\epsfclipon
\usepackage{color}
\usepackage[normalem]{ulem }
\usepackage{soul}
\usepackage{hyperref}
\usepackage{xcolor}
\hypersetup{
    colorlinks,
    linkcolor={red!50!black},
    citecolor={blue!50!black},
    urlcolor={blue!80!black}
}

\definecolor{nred}{RGB}{224,0,0}
\definecolor{nblue}  {RGB}{28,130,185}
\definecolor{dgreen} {RGB}{78,138,21}
\definecolor{norange}{RGB}{230,120,20}

\newcommand{\be}{\begin{equation}}
\newcommand{\ee}{\end{equation}}

\begin{document}

\title{The finite $N$ origin of the Bardeen-Moshe-Bander phenomenon and its extension at $N=\infty$ by singular  fixed points} 

\author{C. Fleming} \email{fleming@lptmc.jussieu.fr}
\affiliation{Sorbonne Universit\'e, CNRS, Laboratoire de Physique Th\'eorique de la Mati\`ere Condens\'ee,  F-75005, Paris, France}
\author{B. Delamotte} \email{delamotte@lptmc.jussieu.fr}
\affiliation{Sorbonne Universit\'e, CNRS, Laboratoire de Physique Th\'eorique de la Mati\`ere Condens\'ee,  F-75005, Paris, France}
\author{S. Yabunaka}\email{yabunaka123@gmail.com}
\affiliation{Department of Physics, Kyushu University, Fukuoka 819-0395, Japan}

\date{\today}

\begin{abstract}

We study the $O(N)$ model in dimension three (3$d$) at large and infinite $N$ and show that the line of fixed points found at $N=\infty$ --the Bardeen-Moshe-Bander (BMB) line-- has an intriguing origin at finite $N$. The large $N$ limit that allows us to find the BMB line must be taken on particular trajectories in the $(d,N)$-plane: $d=3-\alpha/N$ and not at fixed dimension $d=3$. Our study also reveals that the known BMB line is only half of the true line of fixed points, the second half being made of singular fixed points. The potentials of these singular fixed points show a cusp for a finite value of the field and their finite $N$ counterparts a boundary layer.
\end{abstract}

\maketitle

Thirty five years ago, Bardeen, Moshe and Bander (BMB) \cite{bardeen1984spontaneous} discovered an intriguing phenomenon in the euclidean O($N$) models with $(\boldsymbol\varphi^2)^3$ interactions. This has then been intensively studied as a simple toy model for several field theoretical problems \cite{Bardeen_Dilaton,leung1986spontaneous,rosenstein1991dynamical,miransky1985dynamics}. It was shown  that (i) at $d=3$ and $N=\infty$, there exists a finite line of fixed points (FPs), hereafter called the BMB line, that starts at the gaussian FP $G$ and ends at a special point that we call the BMB FP \cite{david1985study,David2,bardeen1984spontaneous,litim2018asymptotic,litim2017fixed}, (ii)  although they are interacting FPs their associated critical exponents are all identical to the gaussian ones except for the BMB FP \cite{david1985study,David2}, (iii) the FPs are twice unstable in the infrared, that is, they are tricritical FPs, and are attractive in the ultraviolet: this is a rare example of  asymptotically safe scalar theories \cite{david1985study,David2,litim2018asymptotic,litim2017fixed}, (iv)  the BMB FP shows a singular effective potential at small field \cite{david1985study,David2,litim2018asymptotic,litim2017fixed}, (v)  because of this singularity, dilatation invariance is spontaneously broken at the BMB FP \cite{AMIT1985371,bardeen1984spontaneous,litim2018asymptotic}, (vi) this spontaneous breakdown implies the existence of a massless dilaton which is a bound state\cite{bardeen1984spontaneous}. 

Many aspects and extensions of the BMB phenomenon have  been explored, some involving fermions and supersymmetry \cite{leung1986spontaneous,bardeen1985spontaneous,rosenstein1991dynamical,gudmundsdottir1985supersymmetric,suzuki1986nontrivial,suzuki1985three}. However, despite intensive efforts, at least two major questions have remained open: Is there a finite $N$ extension of the BMB line of FPs? How can the presence of a singular FP at the end of the BMB line be explained from the large $N$ limit? As for the first question, it is widely believed that the BMB FP has no counterpart at finite $N$ \cite{david1985study,David2,gudmundsdottir1984more,litim2017fixed,matsubara1985large,kessler1985infinite} which would imply that the BMB phenomenon is, at best, a curiosity of the $N=\infty$ case. As for the second question, to the best of our knowledge, no progress has been made these last decades \cite{gudmundsdottir1984more,david1985study,David2,litim2017fixed,matsubara1985large,kessler1985infinite}.

It is the aim of this article to  explain how the  finite $N$ extension of the BMB phenomenon can be obtained and how the correct $N\to\infty$ limit should be taken. One of the subtle points of this limit is that it should not be taken at fixed dimension $d$ but in a correlated way $N\to\infty$ and $d\to 3$. In turn, this study will show that the original BMB line is half the full line of FPs at $N=\infty$ that actually  involves singular FPs in its second half. The reason why the BMB FP is singular (our second question above) will then be natural: it is the contact point between the original BMB line made of regular FPs and the second portion of this line made of singular FPs, hereafter called the singular BMB line.

Let us first review what is known from perturbation theory at finite $N$ and let us show that more information can be extracted from its large $N$ limit than is usually done. Within perturbation theory, the usual tricritical  behavior is described for all $N$ by the massless $(\boldsymbol\varphi^2)^3$ theory whose upper critical dimension is 3  \cite{STEPHEN197389,lewis1978tricritical,pisarski1982fixed}. The tricritical FP, that we call $A_2$, is twice unstable in the infrared --hence its subscript 2-- and emerges from the gaussian FP $G$ below $d=3$. It is found for all $N$ in an $\epsilon=3-d$ expansion \cite{pisarski1982fixed,osborn2018seeking,lewis1978tricritical,Hager_2002}. 

It was shown in \cite{osborn2018seeking} that the four-loop $\beta$-function of the $(\boldsymbol\varphi^2)^3$ coupling $\lambda$ obtained within the $\epsilon$-expansion involves all the leading terms at large $N$. This implies that the flow of this coupling is known at large $N$  and reads \cite{osborn2018seeking}:
\begin{equation}
    N \beta_{\tilde\lambda}=-2\alpha\tilde{\lambda}+12\tilde{\lambda}^2-\pi^2\tilde{\lambda}^3/2+O(1/N)
    \label{beta-function1}
\end{equation}
with the usual rescaling $\lambda=\tilde{\lambda}/N^2$ and $\alpha=\epsilon N$.

In 1982, Pisarski \cite{pisarski1982fixed} noticed that at large $N$ the root of  $\beta_{\tilde\lambda}$  that vanishes for $\epsilon=0$: $\tilde\lambda_-=\tilde\lambda_-^\infty+O(1/N)$ with $\tilde\lambda_-^\infty=4/\pi^2( 3- \sqrt{9-\pi^2\alpha/4})$, exists only on a finite range of dimensions: $d\in[d_{\text{c}}(N),3]$ with $d_{\text{c}}(N)=3-\alpha_{\text{c}}/N$ and $\alpha_{\text{c}}=36/\pi^{2}\simeq 3.65$. Of course, this root corresponds  to the $A_2$ FP. It depends on $N$ and $d$ only through $\alpha$, see Eq. (\ref{beta-function1}). The existence of the line $d_{\text{c}}(N)$ in the  $(d,N)$ plane was interpreted in \cite{osborn2018seeking} as the radius of convergence of the $\epsilon=3-d$ expansion. However, Pisarski interpreted it as the location where $A_2(\alpha)$ had to coincide with another FP to disappear for $d<d_{\text{c}}(N)$ \cite{pisarski1982fixed}. This other FP had to be the second root of $\beta_{\tilde\lambda}$: $\tilde\lambda_+=\tilde\lambda_+^\infty+O(1/N)$ with $\tilde\lambda_+^\infty=4/\pi^2( 3+ \sqrt{9-\pi^2\alpha/4})$ since the two FPs coincide for $d=d_{\text{c}}(N)$ and the $\beta$-function
(\ref{beta-function1}) is exact to order $1/N$. At fixed $\alpha<\alpha_{\text{c}}$, the very existence of this second FP can be questioned because on one hand it is an unavoidable consequence of Eq. (\ref{beta-function1}) but, on the other hand, being nongaussian for $\epsilon=0$, it could seem out of reach of the $\epsilon$-expansion. We call it $\tilde A_3(\alpha)$ because if it indeed exists, it  has three unstable infrared eigendirections as can be checked on Eq. (\ref{beta-function1}). A second problem related to $\tilde A_3(\alpha)$ is that $\tilde\lambda_+ $ is well-defined for all values of $\alpha<\alpha_{\text{c}}$ and therefore this FP seems to exist for all dimensions $d>d_{\text{c}}(N)$, which is doubtful \cite{pisarski1982fixed,Pisarski2}.

To shed some light about what can be reliably extracted from the $1/N$ expansion above, Eq. (\ref{beta-function1}), we first study the $N\to\infty$ limit of $\tilde\lambda_\pm$, that is, of the $A_2(\alpha)$ and $\tilde A_3(\alpha)$ FPs. We notice that in Eq. (\ref{beta-function1}), we first have to set $\beta_{\tilde\lambda}=0$ and then let $N\to\infty$. To get a finite large-$N$ limit, we also have to take the limit $N\to\infty$ at fixed $\alpha=\epsilon N$. Then, we find that each $A_2(\alpha)$ FP with $\alpha\in[0,\alpha_{\text{c}}]$ has a well-defined limit at $N=\infty$ and $d=3$ that we call $A(\alpha)$ whose  coupling is  $\tilde\lambda_-^\infty$  [notice: $A(\alpha=0)=G$]. The line of $A(\alpha)$ FPs continues with the FPs that are  the limits of $\tilde A_3(\alpha)$  with $\alpha\le\alpha_{\text{c}}$, that we call $\tilde A(\alpha)$. Of course, the line made of the $A$ and $\tilde A$  FPs --that we collectively denote by ${\cal A}$-- must be the BMB line and we shall prove below that it is indeed so. This will in turn validate the existence at finite $N$ of the $\tilde A_3(\alpha)$  FPs. A difficulty shows up here because the BMB line is known to be finite since it terminates at the BMB FP. Considering that $\tilde\lambda_+$ increases with decreasing $\alpha$, there must therefore exist some $\alpha=\alpha_{\text{BMB}}$ such that $\tilde A_3$ ceases to exist for $\alpha<\alpha_{\text{BMB}}$. The detailed analysis of the BMB line performed in \cite{david1985study,David2,litim2018asymptotic,litim2017fixed}  shows that the effective  potential of the BMB FP is singular at small field. This suggests that even at finite $N$, the (probable) nonexistence of $\tilde A_3$ for $\alpha<\alpha_{\text{BMB}}$ is not visible on $\tilde\lambda_+(\alpha_{\text{BMB}})$ but requires the computation of the effective potential, that is, requires a functional approach. This is what we are doing below where we confirm the need of a functional approach to fully understand the whole picture of the BMB phenomenon both at finite and infinite $N$. This will in particular explain why and how $\tilde A_3(\alpha)$ disappears at finite $N$ for $\alpha<\alpha_{\text{BMB}}$.

Crucial to our study is the Wilsonian functional and nonperturbative renormalization group (NPRG). As we show below, it does not only provide an efficient means of computing FPs that are out of reach of usual perturbative or $1/N$ approaches but it also allows us to tackle problems that cannot even be formulated in the usual perturbative framework \cite{yabunaka2017surprises,yabunaka2018might,tissier08,tissier10,TissierPhysRevB.74.214419,CanetPhysRevE.93.063101,PhysRevB.69.220408,PhysRevB.64.014412,CanetPhysRevE.84.061128,GredatPhysRevE.89.010102}. Among them are FPs whose potential shows a cusp or a boundary layer. We therefore now give an introduction to the NPRG.

The NPRG is based on the idea of integrating statistical fluctuations step by step \cite{PhysRevB.4.3174}. In its modern version, it
is implemented on the Gibbs free energy $\Gamma$ which is the generating functional of one-particle irreducible correlation functions \cite{wetterich91,wetterich93b,Ellwanger,Morris94,delamotte2012}. A one-parameter family of models indexed by a momentum scale $k$ with partition
functions ${\cal Z}_k$ is
thus defined, such that only the rapid fluctuations, with
wavenumbers $\vert q\vert > k$, are summed over in ${\cal Z}_k$. The decoupling of the slow modes ($\vert q\vert < k$) in ${\cal Z }_k$ is performed by adding to the original hamiltonian $H$ a quadratic (mass-like) term which is nonvanishing
only for these modes:
\begin{equation}
 {\cal Z}_k[\boldsymbol{J}]= \int D\boldsymbol\varphi \exp(-H[\boldsymbol\varphi]-\Delta H_k[\boldsymbol\varphi]+ \boldsymbol{J}\cdot\boldsymbol\varphi)
\end{equation}
 where $\boldsymbol\varphi$ is a $N$-component scalar field, $H$ a general O($N$)-invariant hamiltonian and $\Delta H_k[\boldsymbol\varphi]=\frac{1}{2}\int_q R_k(q^2) \varphi_i(q)\varphi_i(-q)$
 where, for instance, $R_k(q^2)= \bar Z_k (k^2-q^2)\theta(k^2-q^2)$ with $\theta$ the Heaviside function, $\bar Z_k$ the field renormalization,
and $\boldsymbol{J}\cdot\boldsymbol\varphi=\int_x J_i(x) \varphi_i(x)$.
 The $k$-dependent  Gibbs free energy $\Gamma_k[\boldsymbol\phi]$, with $\phi_i=\langle\varphi_i\rangle$,
is defined as  the (slightly modified) Legendre transform of $\log  {\cal Z}_k[\boldsymbol{J}]$:
\begin{equation}
\label{legendre}
 \Gamma_k[\boldsymbol\phi]+\log  {\cal Z}_k[\boldsymbol{J}]= \boldsymbol{J}\cdot\boldsymbol\phi-\frac 1 2 \int_q R_k(q^2) \phi_i(q)\phi_i(-q).
 \end{equation}
The exact RG flow equation of $\Gamma_k$ reads \cite{wetterich93b}:
\begin{equation}
\label{flow}
\partial_t\Gamma_k[\boldsymbol\phi]=\frac 1 2 {\rm Tr} \big[\partial_t R_k(x-y) \left(\Gamma_k^{(2)}+R_k\right)[x,y;\boldsymbol\phi]^{-1}\big]
\end{equation}
where $t=\log(k/\Lambda)$, $\Lambda$ being the ultra-violet cutoff of the model (analogous to the inverse lattice spacing for a lattice model), ${\rm Tr}$ stands for an integral over $x$ and $y$ and a trace over group indices and $\Gamma_k^{(2)}=\Gamma_k^{(2)}[x,y;\boldsymbol\phi]$ is
the matrix of the second functional derivatives of $\Gamma_k[\boldsymbol\phi]$ with respect to $\phi_i(x)$ and $\phi_j(y)$. One can show from Eq. (\ref{legendre}) and from the shape of the regulator that in the limit   $k\to\Lambda$,  no fluctuation is taken into account  and $\Gamma_{k=\Lambda}=H$  while when $k\to 0$, all fluctuations are summed over  and the usual  Gibbs free energy $\Gamma$ is recovered: $\Gamma_{k=0}=\Gamma$.
Thus,  at any  finite scale $k<\Lambda$,  $\Gamma_k$  interpolates between  the hamiltonian $H$ and the Gibbs free energy $\Gamma$.

For the models we are interested in, it is impossible to solve
Eq.~(\ref{flow}) exactly and we therefore have recourse to  approximations. The most appropriate nonperturbative  approximation consists in expanding $\Gamma_k[\boldsymbol\phi]$ in powers of $\nabla\boldsymbol\phi$ \cite{berges2002,delamotte04,canet04,tissier10,tissier08,kloss14,canet16}. At lowest order, called the local potential approximation (LPA), 
$\Gamma_k$ is approximated by:
\begin{equation}
 \Gamma_k^{\text{LPA}}[\boldsymbol\phi] = \int_{x} \left(\frac{1}{2} (\nabla \phi_i)^2 + 
  U_k(\rho)\vphantom{\frac{1}{4}}\right)      
   \label{LPA}
\end{equation}
where $\rho= \phi_i\phi_i/2$.


When at criticality, the system is self-similar and the RG flow reaches a FP which is only reachable when using dimensionless  quantities. We proceed as usual by rescaling
fields and coordinates according to $\tilde x = kx$, and $\tilde{\boldsymbol\phi}(\tilde x) = v_{d}^{-\frac{1}{2}} k^{(2-d)/2} \bar Z_k^{1/2}\boldsymbol\phi(x)$ with 
$v_{d}^{-1}=2^{d-1}d\pi^{d/2}\Gamma(\frac{d}{2})$. The potential is then rescaled according to its canonical dimension $\tilde U_k(\tilde{\rho})=v_{d}^{-1} k^{-d} U_k(\rho)$. Notice that there is no field renormalization at the LPA level, that is, $\bar Z_k=1$, which implies that the rescaling of $\boldsymbol\phi$ is performed according to its canonical dimension only and the anomalous dimension at the fixed point is vanishing.

In practice, computing the flow of the potential $U_k$ requires several steps. First, the potential is defined by: $U_k(\boldsymbol\phi)=\Gamma_k[\boldsymbol\phi]$ where $\boldsymbol\phi$ is a constant field. Then, the flow of $U_k(\boldsymbol\phi)$ is obtained by acting with $\partial_t$ on both sides of the above definition of $U_k$ and by using Eq. (\ref{flow}). Finally, $\Gamma_k^{(2)}$ in the right hand side of Eq. (\ref{flow})  is computed from the LPA ansatz, Eq. (\ref{LPA}).



\begin{figure}[t]
\begin{centering}
\includegraphics[width=0.7\columnwidth]{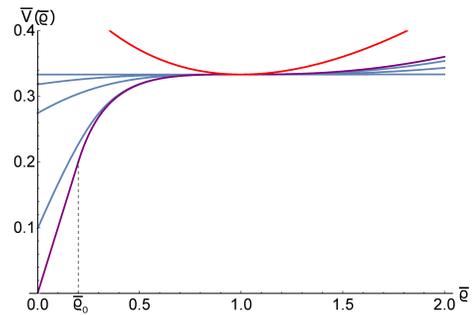} 
\par\end{centering}
\caption{Potentials $\bar V(\bar\varrho)$ of the tricritical FPs ${\cal A}(\tau)$  of the BMB line (blue)  together with the Wilson-Fisher FP (red). The BMB FP is the endpoint of the BMB line (purple). All these potentials are given by Eq. (\ref{solrho}) (in the Wilson-Polchinski version of the LPA flow). The gaussian FP $G$ corresponds to the horizontal line. The BMB FP potential shows a discontinuity in its second derivative at $\bar\varrho_0$.
}
\label{V(rho)}
\end{figure}

It is very convenient for the following to make a change of variables that simplifies considerably the study of the large $N$ limit. This change of variables is equivalent to working with the Wilson-Polchinski flow instead of the flow of $\Gamma_k$, hereafter called the Wetterich flow.  Following Ref. \cite{Morris}, we define:
 $\tilde{V}(\tilde{\varrho})=\tilde{U}(\tilde{\rho})+\left(\tilde{\phi}_i-\tilde{\Phi}_i \right)^2/2$ with $\tilde{\varrho}=\tilde{\Phi}_i \tilde{\Phi}_i/2=\tilde{\Phi}^2 /2$ and $\tilde{\phi}_i-\tilde{\Phi}_i=-\tilde{\Phi}_i \tilde{V}'(\tilde{\varrho})=-\tilde{\phi}_i \tilde{U}'(\tilde{\rho})$.  It is convenient to rescale $\tilde{\varrho}$ and $\tilde{V}(\tilde{\varrho})$ as usual:   $\bar\varrho=\tilde{\varrho}/{N}$, $\bar V=\tilde{V}/N$.  The FP equation for $\bar V(\bar\varrho)$  thus reads \cite{litim2018asymptotic,Morris,yabunaka2018might} 
\begin{equation}
 0=1-d\,\bar V+(d-2)\bar\varrho \bar V'+2\bar\varrho{\bar V'}{}^2-\bar V'-\frac{2}{N}\bar\varrho\,\bar{V}''
\label{flow-LPA-WP-essai}
\end{equation}
where the primes mean derivatives with respect to $\bar\varrho$. Here again, the usual  $N\to\infty$ limit consists in assuming that $\bar V(\bar\varrho)$ is regular for all $\bar\varrho$ and thus in discarding the last term in Eq. (\ref{flow-LPA-WP-essai}) because of its $1/N$ prefactor.  In $d=3$ and $N=\infty$, there are infinitely many solutions to Eq.~(\ref{flow-LPA-WP-essai}) in which the last term has been discarded \cite{bardeen1984spontaneous,Omid,litim2018asymptotic,litim2017fixed}. They are given by the following implicit expression valid for the physical solutions of interest here:
\begin{equation}
\bar{\varrho}_{\pm}=1+\frac{\bar{V}'\left(\frac{5}{2}-\bar{V}'\right)}{\left(1-\bar{V}'\right)^{2}}+\frac{\frac{3}{2}\arcsin\sqrt{\bar{V}'}\pm \sqrt{2/\tau}}{\left(\bar{V}'\right)^{-1/2}\left(1-\bar{V}'\right)^{5/2}}
\label{solrho}
\end{equation}
 where $\bar{\varrho}_{+}\left(\bar{V}'\right)$ and $\bar{\varrho}_{-}\left(\bar{V}'\right)$ correspond to the two branches $\bar{\varrho}>1$ and $\bar{\varrho}<1$ respectively, and $\tau$ is an integration constant. A detailed analysis of Eq. (\ref{solrho}) shows that (i) the gaussian FP $G$ for which $\bar V'(\bar\varrho)=0$ is obtained for $\tau=0$, (ii) a well-defined solution $\bar V(\bar\varrho)$ exists for all $\tau\in[0,\tau_{\text{ BMB}}=32/(3\pi)^2]$ which therefore corresponds to the BMB line of FPs, denoted here by ${\cal A}(\tau)$, with the BMB FP being the endpoint obtained for $\tau=\tau_{\text{ BMB}}$ as in \cite{bardeen1984spontaneous,david1985study,David2,litim2018asymptotic,litim2017fixed}, (iii) for $\tau>\tau_{\text{ BMB}}$  the solutions of Eq. (\ref{solrho}) are not defined on the whole interval $\bar\varrho\in[0,\infty[$ \cite{litim2018asymptotic}, (iv)  an isolated solution exists for $\sqrt{2/\tau}=0$ which corresponds to the Wilson-Fisher FP associated with the usual second order phase transition of the O($N=\infty$) model (an analytic continuation is needed when $V'<0$).  All the FPs corresponding to a value of  $\tau\in[0,\tau_{\text{ BMB}}[$ are twice unstable in the infrared and are tricritical.  
We plot their potential in Fig. \ref{V(rho)}. One observes that, for all $\tau<\tau_{\text{BMB}}$, the  FP potentials $\bar{V}_\tau(\bar{\varrho})$ along the BMB line are regular for all values of the field. Approaching $\tau_{\text{BMB}}$, the FP potentials approach a limiting shape which shows a singularity at a value $\bar{\varrho}_0$  in its second derivative, see Fig. \ref{V(rho)} and a detailed description below
 (see also Fig. 1 of the Supplemental Material)
\footnote{ Notice that the linear part of the BMB FP potential corresponding to $\bar\varrho\in[0,\bar\varrho_0]$, see Fig. \ref{V(rho)}, can be replaced by a smooth analytic continuation of the potential obtained for $\bar\varrho>\bar\varrho_0$. Considering either solution has no physical consequence because in both cases, the interval $\bar\varrho\in[\bar\varrho_0,\infty[$ is entirely mapped onto $\bar\rho\ge0$ in the  Wetterich version of the flow.}. Notice that in the  Wetterich version of the flow \cite{wetterich93b,berges2002}, the potential of the BMB FP shows a singularity at vanishing field \cite{litim2017fixed}.

Let us now look for the finite $N$ origin of the BMB line within our functional framework. Just as in perturbation theory, we take the limit $N\to\infty$ and $d\to3$ at fixed $\alpha=\epsilon N$. Our aim is to show that to each FP  ${\cal A}(\tau)$ with $\tau\in[0,\tau_{\text{ BMB}}]$ on the BMB line, there is one FP at finite $N$, either $A_2(\alpha)$ or $\tilde A_3(\alpha)$, that converges to ${\cal A}(\tau)$ when $N\to\infty$. The problem is therefore to relate admissible values of $\tau$, that is, values for which a FP on the BMB line exists, to admissible values of $\alpha$ where $A_2(\alpha)$ or $\tilde A_3(\alpha)$ exist.

Within the LPA, the proof goes as follows. We assume that at large $N$, the FP potentials can be expanded as: 
\begin{equation}
\bar V_{\alpha,N}(\bar\varrho) =\bar V_{\alpha,N=\infty}(\bar\varrho)+\bar V_{1,\alpha}(\bar\varrho)/N +O(1/N^2).
\label{expansion}
\end{equation}
We assume that $\bar V_{\alpha,N}(\bar\varrho)$, $\bar V_{1,\alpha}(\bar\varrho)$ and $\bar V_{\alpha,N=\infty}(\bar\varrho)$ are regular functions of $\bar\varrho$. As such, $\bar V_{\alpha,N=\infty}(\bar\varrho)$ must  be the potential of one the FPs on the BMB line. It must therefore correspond to a solution of Eq. (\ref{solrho}) with a definite value of $\tau\in[0,\tau_{\text{ BMB}}]$: $\bar V_{\alpha,N=\infty}(\bar\varrho)=\bar V_{\tau}(\bar\varrho)$. We therefore conclude that the regularity of $\bar{V}_{1,\alpha}(\bar\varrho)$ together with Eqs. (\ref{flow-LPA-WP-essai}) and (\ref{expansion}) determines the relation between $\tau$ and $\alpha$.

It is particularly convenient to impose the analyticity of $\bar{V}_{1,\alpha}(\bar\varrho)$ at $\bar\varrho=1$ because all FPs of the BMB line show an inflection point for this value of  $\bar\varrho$ as can be seen on Fig. \ref{V(rho)}.
Generically, a nonanalytic logarithmic behavior shows up at this point when substituting Eq. (\ref{expansion}) into Eq. (\ref{flow-LPA-WP-essai}). Requiring that its prefactor vanishes imposes (see Section 3 of the Supplemental Material):
\begin{equation}
\text{\ensuremath{\alpha}}-36\tau+96\tau^2=0.
\label{eq:alpha1}
\end{equation}
This equation has two solutions  $\tau_1(\alpha)$ and $\tau_2(\alpha)$ that we choose such that $\tau_1(\alpha)\le\tau_2(\alpha)$ for all $\alpha$, see Fig. 2 of the Supplemental Material. This implies that to each value of $\alpha$, that is, to each point on the hyperbola $d=3-\alpha/N$, correspond two FPs on the BMB line that, as in  perturbation theory,  are $A=A(\tau_1(\alpha))$ and $\tilde A=\tilde A(\tau_2(\alpha))$. According to Eq. (\ref{expansion}), for any value of $\alpha$, these FPs must be the limits of two different  FPs existing at finite $N$: They are nothing but the  $A_2(\alpha)$ and $\tilde A_3(\alpha)$ FPs found perturbatively from Eq. (\ref{beta-function1}) with, by definition, $A_2(\alpha)\to A(\tau_1(\alpha))$ and $\tilde A_3(\alpha)\to \tilde A(\tau_2(\alpha))$ when $N\to\infty$.

Let us first notice that, as expected,  $A(\tau=0)=G$ since   $A_2(\alpha=0)=G$, $\forall N$. For finite $\alpha$, the $A_2(\alpha)$ FPs are continuously related to $G$ by continuously decreasing $d$ at fixed $N$ and their limit at $N=\infty$ must therefore also be continuously related to $G$ on the BMB line. This is the $A(\tau_1(\alpha))$ branch of this line. This branch meets the other one for  $\tau_1(\alpha)=\tau_2(\alpha)\equiv \tau_{\text{c}}$, that is, $A(\tau_{\text{c}})=\tilde A(\tau_{\text{c}})$. At finite $N$, this indicates that $A_2(\alpha)=\tilde A_3(\alpha)$, which, by definition, occurs for $\alpha=\alpha_{\text{c}}$. Within the LPA, we  find  $\alpha_{\text{c}}^{\text{LPA}}=27/8= 3.375$ instead of the exact result $36/\pi^2\simeq3.65$ obtained from Eq.~(\ref{beta-function1}). 
For  values of $\tau$ larger than $\tau_{\text{c}}$, that is,  $\tau=\tau_2(\alpha)\in[\tau_{\text{c}},\tau_{\text{BMB}}]$, the $\tilde A(\tau)$ FPs on the BMB line are the limits of $\tilde A_3(\alpha)$. Using Eq.(\ref{eq:alpha1}) we find for $\tau$ that its upper bound $\tau_{\text{BMB}}$  translates into a lower bound on $\alpha$: $\alpha_{\text{BMB}}=\alpha(\tau_{\text{BMB}})$. At the LPA, we find  from Eq.~(\ref{eq:alpha1}): $\alpha_{\text{BMB}}^{\text{LPA}}\simeq0.51$.

The first order in the $1/N$ expansion performed in Eq.~(\ref{expansion}) does not allow us to fully understand how $\tilde A_3(\alpha)$ disappears at finite $N$ for $\alpha<\alpha_{\text{BMB}}$ and we therefore have had recourse to a numerical integration of the flow.
We have checked numerically at finite and large $N$ (typically $N>75$) by directly integrating Eq. (\ref{flow-LPA-WP-essai}) that all the results described previously are indeed correct within the LPA. More precisely, we have checked the following:  existence of $A_2$ for all $N$ that emerges from $G$ below $d=3$, collapse of $A_2$ with another, three times unstable FP $\tilde A_3$ on the line $d_{\text{c}}(N)$ \cite{yabunaka2017surprises}, existence of $\tilde A_3$ on a finite interval  $\alpha_{\text{BMB}}\le\alpha\le\alpha_{\text{c}}$ only, existence of  well-defined limits of $A_2(\alpha)$ and $\tilde A_3(\alpha)$ when $N\to\infty$ given by the potentials of $A(\tau_1(\alpha))$ and $\tilde A(\tau_2(\alpha))$ as found in Eq.~(\ref{solrho}).

 Our numerical analysis of Eq. (\ref{flow-LPA-WP-essai}) raises two paradoxes of the $1/N$ analysis above. The first one is related to the question: How is it possible that $\tilde A_3$ disappears at finite $N$ for $\alpha<\alpha_{\text{BMB}}$? Usually, a FP disappears by colliding with another one. We have numerically found that this is again what happens here: there is indeed another FP, different from $A_2$, with which $\tilde A_3$ collides at $\alpha=\alpha_{\text{BMB}}$. We call it $S\tilde{A}_4$ where the meaning of the $S$ will be clarified in the following.  A first paradox appears here: $S\tilde{A}_4$ has no counterpart at $N=\infty$ in the BMB line although the potentials of Eq. (\ref{solrho}) are supposed to constitute the complete set of solutions of Eq. (\ref{flow-LPA-WP-essai}) at $N=\infty$. The second question is: On which range of values of $\alpha$ does $S\tilde{A}_4$ exist? We have  found that at fixed and large $N$, it also exists on a finite interval of dimensions $d$ with $d<3-\alpha_{\text{BMB}}/N$ and that it collides  with yet another FP, that we call $SA_3$ \footnote{ In \cite{yabunaka2018might}, $SA_3$ was called $C_3$.}. We have numerically found  at very large $N$ that this collision between  $S\tilde{A}_4$ and $SA_3$ seems to occur at the same value $\alpha=\alpha_{\text{c}}$ where $A_2$ collides with $\tilde A_3$. 
 The second paradox is that again there is no counterpart of $SA_3$ at $N=\infty$ and $d=3$ in the potentials of Eq. (\ref{solrho}). 
 To summarize, we have numerically found for finite and very large values of $N$ that for $\alpha_{\text{BMB}}<\alpha<\alpha_{\text{c}}$ four FPs exist: $A_2, \tilde A_3, S\tilde{A}_4, SA_3$. The remarkable values of $\alpha$ correspond to the collision between these FPs: $A_2(\alpha_{\text{c}})= \tilde A_3(\alpha_{\text{c}})$, $\tilde A_3(\alpha_{\text{BMB}})=S\tilde{A}_4(\alpha_{\text{BMB}})$, $S\tilde{A}_4(\alpha_{\text{c}})= SA_3(\alpha_{\text{c}})$ and we recall that $A_2(0)= G$. 

The two paradoxes described above are solved by realizing that the potentials of both $S\tilde{A}_4$ and $SA_3$ become singular in the $N\to\infty$ limit: Strictly speaking, since they show a singularity at $N=\infty$, they are not solutions of Eq. (\ref{flow-LPA-WP-essai}) in which the last term has been discarded and therefore do not belong to the set of solutions given in Eq. (\ref{solrho}). We show below both at finite and infinite $N$ how these FPs can be fully analytically characterized.

\begin{figure}
\begin{centering}
\includegraphics[width=0.7\columnwidth]{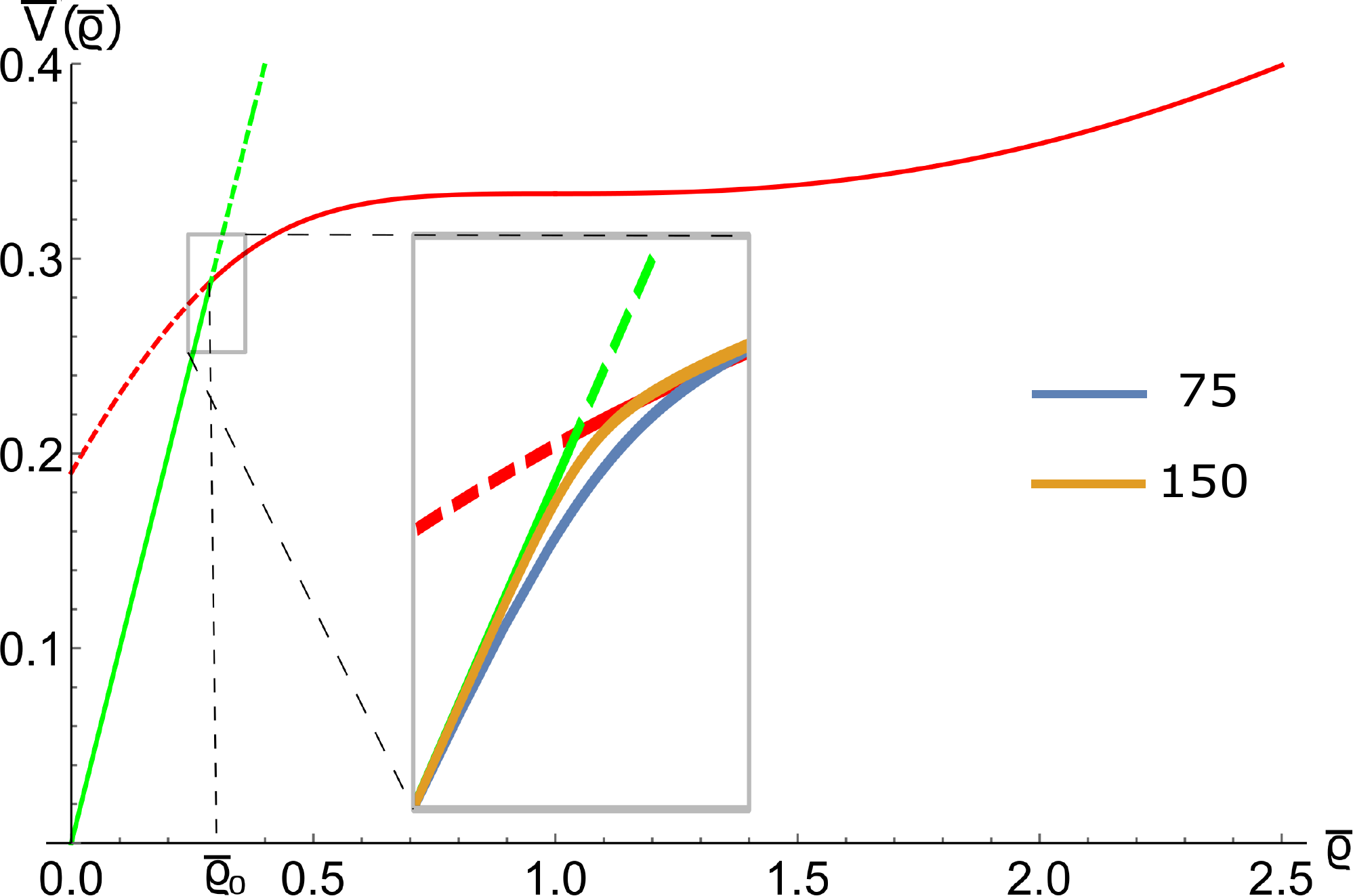}
\par\end{centering}
\caption{$N=\infty$ and $d=3$: Singular potential of $S{\cal A}(\tau=0.33)$  from the potential of ${\cal A}(\tau=0.33)$  given by the red and dashed red curves, Eq. (\ref{solrho}). The green and dashed green curves show $\bar{V}(\bar\varrho)=\bar\varrho$. The potential of $S{\cal A}(\tau=0.33)$ is made of the plain green and red curves that meet at $\bar\varrho_0(\tau=0.33)$. Inset:  zoom of the region around the cusp and its rounding at finite $N$ within the boundary layer.
}
\label{sing-pot-polchinski}
\end{figure}

Let us  start by the $N=\infty$ case. Since the potentials we are interested in are singular, we must enlarge the space of functions in which Eq. (\ref{flow-LPA-WP-essai}) is solved at $N=\infty$. We consider in the following only solutions that show a cusp at an isolated value of $\bar\varrho$: They will be obtained by matching together solutions of Eq. (\ref{flow-LPA-WP-essai}), with the $1/N$ term discarded, that are piecewise well-defined.

We first notice that the second derivative of the potential of the BMB FP shows a discontinuity at $\bar\varrho_0$, as already mentioned above, see Fig. \ref{V(rho)} and Fig. 1 of the Supplemental Material. This potential is indeed made of two parts: a linear part $\bar{V}(\bar\varrho)=\bar\varrho$ up to the value $\bar\varrho_0$ and a part for $\bar\varrho>\bar\varrho_0$ where $\bar{V}''(\bar\varrho)\neq0$. The value of $\bar\varrho_0$ is determined by the continuity of $\bar{V}'(\bar\varrho_0)$, see Fig. \ref{V(rho)}. From the construction of the BMB FP potential described above, it is simple to build a ``singular copy" of the BMB line: Take any FP on this line, either $\tilde{A}(\tau)$ or $A(\tau)$, and, as shown in Fig. \ref{sing-pot-polchinski}, replace the small $\bar\varrho$ part of the potential  by the straight line $\bar{V}(\bar\varrho)=\bar\varrho$ up to the point $\bar\varrho_0(\tau)$ where the two curves meet. The result is obviously a solution of Eq. (\ref{flow-LPA-WP-essai}) at $N=\infty$ for all  $\bar\varrho\neq\bar\varrho_0(\tau)$. It is therefore a FP potential with the peculiarity that it shows a cusp at $\bar\varrho_0(\tau)$. We call $S\tilde{A}(\tau)$ and $SA(\tau)$ the resulting FPs which are generically denoted  by $S{\cal A}(\tau)$, the $S$ meaning singular. We thus discover that the usual BMB line is actually only half of the true line of FPs at $N=\infty$. In the construction above, the BMB FP plays a pivotal role since all singular FPs are obtained by continuously deforming its potential. 

We now show that the singular FPs $S\tilde{A}(\tau)$ and $SA(\tau)$ are the limits at $N=\infty$ and $d=3$ of the $S\tilde{A}_4(\alpha)$ and $SA_3(\alpha)$ FPs  in much the same way as $A(\tau_1(\alpha))$ and $\tilde A(\tau_2(\alpha))$ are the limits of $A_2(\alpha)$ and $\tilde A_3(\alpha)$. The construction of the potentials of the finite and large $N$ extensions
of $S{\cal A}(\tau)$ relies on the idea that the cusp of these potentials builds up as $N$ increases, that is, the cusp is smoothened at finite $N$ and shows up only in the limit $N\to\infty$. 
In other words, at finite $N$, there should exist a boundary layer around the point $\bar\varrho_0(\tau)$, such that inside the layer the potential varies smoothly -- but abruptly -- in order to connect the linear part of the potential for $\bar\varrho<\bar\varrho_0(\tau)$ to the nontrivial part of the potential for $\bar\varrho>\bar\varrho_0(\tau)$. Moreover, the boundary layer should be sufficiently thin such that $\bar{V}''$  varies as $N$  at large $N$ in such a way that inside the layer it compensates the $1/N$ factor in front of it in Eq. (\ref{flow-LPA-WP-essai}). This is achieved by a layer of typical width $1/N$. In this case,  all the terms of Eq. (\ref{flow-LPA-WP-essai}), including the last one, must be retained in the large $N$ limit because they are all of the same order in $N$. 

Finding the boundary layer is easier done with $\bar{V}'(\bar\varrho)$  rather than with $\bar{V}(\bar\varrho)$ (see section 4 of the Supplemental Material for details). We define the scaled variable: $\tilde{\varrho}=N\left(\bar{\varrho}-\bar{\varrho}_{0}\right)$ inside the layer. Then, we find that in terms of this variable,  $F(\tilde{\varrho})=\bar{V}'(\bar\varrho)$  satisfies inside the boundary layer and at leading order in $1/N$: $0=1-3\,\bar{V}(\bar{\varrho}_0)+\bar\varrho_0 F +2\bar\varrho_0\left({F}^2-F'\right) - F.$
The solution of this equation reads: $F(\tilde{\varrho})=V_{1}-V_{2}\tanh\left(V_{2}\,\tilde{\varrho}\right)$ with $2V_{i}=V'\left(\bar{\varrho_{0}}^{-}\right)\pm V'\left(\bar{\varrho_{0}}^{+}\right)$
where the plus sign goes with $i=1$ and the minus sign with $i=2$.
It is then straightforward to show that this solution connects smoothly the two values $\bar{V}'(\bar{\varrho}_0^{\,-})$ and $\bar{V}'(\bar{\varrho}_0^+)$ across the boundary layer, as expected. Notice that the existence of a boundary layer cannot be found by usual perturbative means that assume that the rescaling of $\varrho$ and of the potential by a       factor of $N$  --which leads to the usual scaling of $\lambda$ in Eq.~(\ref{beta-function1})-- does not depend on the value of $\varrho$. 

Now that we have shown that the potentials of the $S{\cal A}(\tau)$ FPs have a possible extension at finite $N$, we have to study on which interval of dimensions  $d=3-\alpha/N$ these FPs exist. The crucial remark here is that the potential of $S{\cal A}(\tau)$  is identical to that  of ${\cal A}(\tau)$ for $\bar\varrho>\bar\varrho_0(\tau)$ and in particular at $\bar\varrho=1$. For these values of $\bar\varrho$, the argument used to connect the FPs $A_2(\alpha)$ and $\tilde A_3(\alpha)$ found at finite $N$ to the FPs $A(\tau_1)$ and $\tilde{A}(\tau_2)$ at $N=\infty$ can be repeated for the singular FPs. It yields the same conclusion: There exists two FPs at finite $N$ whose limits are $SA(\tau_1)$ and $S\tilde{A}(\tau_2)$. These FPs are of course those that were found numerically, that is, $S\tilde{A}_4(\alpha)$ and $SA_3(\alpha)$ and  the relation between  $\alpha$ and $\tau$ is again given by Eq. (\ref{eq:alpha1}). At asymptotically large $N$, the two FPs $SA_3(\alpha)$ and $S\tilde{A}_4(\alpha)$ must therefore collide on the same line $d_{\text{c}}(N)=3-\alpha_{\text{c}}/N$ as $A_2(\alpha)$ and $\tilde A_3(\alpha)$. 
Since for $\tau=\tau_{\text{BMB}}$, $S\tilde{A}=\tilde{A}$ = BMB FP, we must have  at large $N$: $S\tilde{A}_4(\alpha_{\text{BMB}})=\tilde{A}_3(\alpha_{\text{BMB}})$. Therefore, at large $N$, $S\tilde{A}_4(\alpha)$  exists only on the interval: $\alpha\in[\alpha_{\text{BMB}}, \alpha_{\text{c}}]$. All these results prove analytically what was empirically found in our numerical analysis. [The numerical analysis makes easy the determination of the number of infrared unstable directions for each FP]. As for $SA_3$, we have been able to follow it up to dimensions significantly larger than 3  where it is no longer related to the BMB phenomenon. A full study of this FP at finite and infinite $N$, together with other nontrivial ones, will be given in a forthcoming publication.

Finally, let us notice that the exact value of $\alpha_{\text{BMB}}$ can be computed from the $N=\infty$ analysis. The effective potentials of the FPs along the BMB line are all regular at small $\tilde\lambda$ \cite{david1985study,David2, litim2017fixed} and it is only from the BMB FP that the potentials start showing a singularity at small fields. This has been shown to occur for $\tilde\lambda_+^\infty=2=\lambda_{\text{BMB}}$ \cite{bardeen1984spontaneous,david1985study,David2,osborn2018seeking}.  Provided that Eq.~(\ref{beta-function1}) is exact at order $1/N$, the corresponding exact value of $\alpha$  is  $\alpha_{\text{BMB}}=12-\pi^2\simeq2.13$. Let us notice that whereas the LPA value of $\alpha_{\text{c}}$ is not too far from the exact value -- 3.375 instead of 3.65 -- the LPA value of $\alpha_{\text{BMB}}$ is quantitatively off by a factor 4: It is 0.51 instead of 2.13. 

To conclude, we have found  at $N=\infty$ and $d=3$ that the usual, regular, BMB line represents only half of the full BMB line which is made of both  regular and singular FPs. In the Wilson-Polchinski RG framework, the singular branch of this line consists of FPs whose potential $\bar V(\bar\varrho)$ starts at small field by a linear part followed at larger fields by a regular tricritical potential. At the points $\bar\varrho_0$ where these two parts connect, these singular FP potentials  show a cusp. The BMB FP is the pivotal point between the regular branch of the BMB line and the singular branch. All FPs of the BMB line, either regular or singular, are the limits of FPs existing at finite $N$ with the subtlety that the $N\to\infty$ limit should be taken together with $d\to3$, letting $\alpha=(3-d)N$ fixed.More precisely, the regular branch of the BMB line is obtained as the limit of two sets of FPs, $A_2(\alpha)$ and $\tilde{A}_3(\alpha)$. The singular branch is the limit of two other sets of FPs, namely $SA_3(\alpha)$ and $S\tilde{A}_4(\alpha)$, whose potentials show a boundary layer at finite $N$ that becomes a cusp at $N=\infty$.
At large $N$, all these FPs exist on finite intervals of $d$.

Our analysis of the  BMB phenomenon raises several questions that we now list. First, the value of $\alpha_{\text{BMB}}$ found at the LPA is quantitatively rather poor compared to other quantities determined at the same order \cite{canet04,CanetPhysRevLett.95.100601}. A study at the next order of the derivative expansion shows that it improves as well as $\alpha_{\text{c}}$ and that it can be further improved by studying the dependence of these numbers on the choice of regulator function $R_k(q)$ \cite{canet03,canet05,leonard15,JakubczykPhysRevE.90.062105,Balog2019}: This will be the subject of a forthcoming publication \cite{fleming}. Another challenge is to follow all FPs in the whole $(d,N)$ plane and more precisely at moderate and small $N$. This study has been partly done in \cite{yabunaka2018might} and will be fully clarified in a forthcoming publication. It would also be interesting to know whether the same BMB phenomenon occurs for all multicritical FPs of the O($N$) models around their respective upper critical dimensions  and whether it exists  generically for models different from the O($N$) models \cite{yabunaka2017surprises}.

S. Y. was supported by Grant-in-Aid for Young Scientists (B) (15K17737 and 18K13516).

\bibliographystyle{apsrev4-1} 

\bibliography{biblio}

\begin{thebibliography}{56}%
\makeatletter
\providecommand \@ifxundefined [1]{%
 \@ifx{#1\undefined}
}%
\providecommand \@ifnum [1]{%
 \ifnum #1\expandafter \@firstoftwo
 \else \expandafter \@secondoftwo
 \fi
}%
\providecommand \@ifx [1]{%
 \ifx #1\expandafter \@firstoftwo
 \else \expandafter \@secondoftwo
 \fi
}%
\providecommand \natexlab [1]{#1}%
\providecommand \enquote  [1]{``#1''}%
\providecommand \bibnamefont  [1]{#1}%
\providecommand \bibfnamefont [1]{#1}%
\providecommand \citenamefont [1]{#1}%
\providecommand \href@noop [0]{\@secondoftwo}%
\providecommand \href [0]{\begingroup \@sanitize@url \@href}%
\providecommand \@href[1]{\@@startlink{#1}\@@href}%
\providecommand \@@href[1]{\endgroup#1\@@endlink}%
\providecommand \@sanitize@url [0]{\catcode `\\12\catcode `\$12\catcode
  `\&12\catcode `\#12\catcode `\^12\catcode `\_12\catcode `\%12\relax}%
\providecommand \@@startlink[1]{}%
\providecommand \@@endlink[0]{}%
\providecommand \url  [0]{\begingroup\@sanitize@url \@url }%
\providecommand \@url [1]{\endgroup\@href {#1}{\urlprefix }}%
\providecommand \urlprefix  [0]{URL }%
\providecommand \Eprint [0]{\href }%
\providecommand \doibase [0]{http://dx.doi.org/}%
\providecommand \selectlanguage [0]{\@gobble}%
\providecommand \bibinfo  [0]{\@secondoftwo}%
\providecommand \bibfield  [0]{\@secondoftwo}%
\providecommand \translation [1]{[#1]}%
\providecommand \BibitemOpen [0]{}%
\providecommand \bibitemStop [0]{}%
\providecommand \bibitemNoStop [0]{.\EOS\space}%
\providecommand \EOS [0]{\spacefactor3000\relax}%
\providecommand \BibitemShut  [1]{\csname bibitem#1\endcsname}%
\let\auto@bib@innerbib\@empty
\bibitem [{\citenamefont {Bardeen}\ \emph {et~al.}(1984)\citenamefont
  {Bardeen}, \citenamefont {Moshe},\ and\ \citenamefont
  {Bander}}]{bardeen1984spontaneous}%
  \BibitemOpen
  \bibfield  {author} {\bibinfo {author} {\bibfnamefont {W.~A.}\ \bibnamefont
  {Bardeen}}, \bibinfo {author} {\bibfnamefont {M.}~\bibnamefont {Moshe}}, \
  and\ \bibinfo {author} {\bibfnamefont {M.}~\bibnamefont {Bander}},\ }\href
  {\doibase 10.1103/PhysRevLett.52.1188} {\bibfield  {journal} {\bibinfo
  {journal} {Phys. Rev. Lett.}\ }\textbf {\bibinfo {volume} {52}},\ \bibinfo
  {pages} {1188} (\bibinfo {year} {1984})}\BibitemShut {NoStop}%
\bibitem [{\citenamefont {Bardeen}\ \emph {et~al.}(1986)\citenamefont
  {Bardeen}, \citenamefont {Leung},\ and\ \citenamefont
  {Love}}]{Bardeen_Dilaton}%
  \BibitemOpen
  \bibfield  {author} {\bibinfo {author} {\bibfnamefont {W.~A.}\ \bibnamefont
  {Bardeen}}, \bibinfo {author} {\bibfnamefont {C.~N.}\ \bibnamefont {Leung}},
  \ and\ \bibinfo {author} {\bibfnamefont {S.~T.}\ \bibnamefont {Love}},\
  }\href {\doibase 10.1103/PhysRevLett.56.1230} {\bibfield  {journal} {\bibinfo
   {journal} {Phys. Rev. Lett.}\ }\textbf {\bibinfo {volume} {56}},\ \bibinfo
  {pages} {1230} (\bibinfo {year} {1986})}\BibitemShut {NoStop}%
\bibitem [{\citenamefont {Leung}\ \emph {et~al.}(1986)\citenamefont {Leung},
  \citenamefont {Love},\ and\ \citenamefont {Bardeen}}]{leung1986spontaneous}%
  \BibitemOpen
  \bibfield  {author} {\bibinfo {author} {\bibfnamefont {C.}~\bibnamefont
  {Leung}}, \bibinfo {author} {\bibfnamefont {S.}~\bibnamefont {Love}}, \ and\
  \bibinfo {author} {\bibfnamefont {W.~A.}\ \bibnamefont {Bardeen}},\ }\href
  {\doibase https://doi.org/10.1016/0550-3213(86)90382-2} {\bibfield  {journal}
  {\bibinfo  {journal} {Nucl. Phys. B}\ }\textbf {\bibinfo {volume} {273}},\
  \bibinfo {pages} {649 } (\bibinfo {year} {1986})}\BibitemShut {NoStop}%
\bibitem [{\citenamefont {Rosenstein}\ \emph {et~al.}(1991)\citenamefont
  {Rosenstein}, \citenamefont {Warr},\ and\ \citenamefont
  {Park}}]{rosenstein1991dynamical}%
  \BibitemOpen
  \bibfield  {author} {\bibinfo {author} {\bibfnamefont {B.}~\bibnamefont
  {Rosenstein}}, \bibinfo {author} {\bibfnamefont {B.~J.}\ \bibnamefont
  {Warr}}, \ and\ \bibinfo {author} {\bibfnamefont {S.~H.}\ \bibnamefont
  {Park}},\ }\href {\doibase https://doi.org/10.1016/0370-1573(91)90129-A}
  {\bibfield  {journal} {\bibinfo  {journal} {Phys. Rep.}\ }\textbf {\bibinfo
  {volume} {205}},\ \bibinfo {pages} {59 } (\bibinfo {year}
  {1991})}\BibitemShut {NoStop}%
\bibitem [{\citenamefont {Miransky}(1985)}]{miransky1985dynamics}%
  \BibitemOpen
  \bibfield  {author} {\bibinfo {author} {\bibfnamefont {V.}~\bibnamefont
  {Miransky}},\ }\href@noop {} {\bibfield  {journal} {\bibinfo  {journal}
  {Nuovo Cim.}\ }\textbf {\bibinfo {volume} {90}},\ \bibinfo {pages} {149}
  (\bibinfo {year} {1985})}\BibitemShut {NoStop}%
\bibitem [{\citenamefont {David}\ \emph {et~al.}(1985)\citenamefont {David},
  \citenamefont {Kessler},\ and\ \citenamefont {Neuberger}}]{david1985study}%
  \BibitemOpen
  \bibfield  {author} {\bibinfo {author} {\bibfnamefont {F.}~\bibnamefont
  {David}}, \bibinfo {author} {\bibfnamefont {D.~A.}\ \bibnamefont {Kessler}},
  \ and\ \bibinfo {author} {\bibfnamefont {H.}~\bibnamefont {Neuberger}},\
  }\href@noop {} {\bibfield  {journal} {\bibinfo  {journal} {Nucl. Phys. B}\
  }\textbf {\bibinfo {volume} {257}},\ \bibinfo {pages} {695} (\bibinfo {year}
  {1985})}\BibitemShut {NoStop}%
\bibitem [{\citenamefont {David}\ \emph {et~al.}(1984)\citenamefont {David},
  \citenamefont {Kessler},\ and\ \citenamefont {Neuberger}}]{David2}%
  \BibitemOpen
  \bibfield  {author} {\bibinfo {author} {\bibfnamefont {F.}~\bibnamefont
  {David}}, \bibinfo {author} {\bibfnamefont {D.~A.}\ \bibnamefont {Kessler}},
  \ and\ \bibinfo {author} {\bibfnamefont {H.}~\bibnamefont {Neuberger}},\
  }\href {\doibase 10.1103/PhysRevLett.53.2071} {\bibfield  {journal} {\bibinfo
   {journal} {Phys. Rev. Lett.}\ }\textbf {\bibinfo {volume} {53}},\ \bibinfo
  {pages} {2071} (\bibinfo {year} {1984})}\BibitemShut {NoStop}%
\bibitem [{\citenamefont {Litim}\ and\ \citenamefont
  {Trott}(2018)}]{litim2018asymptotic}%
  \BibitemOpen
  \bibfield  {author} {\bibinfo {author} {\bibfnamefont {D.~F.}\ \bibnamefont
  {Litim}}\ and\ \bibinfo {author} {\bibfnamefont {M.~J.}\ \bibnamefont
  {Trott}},\ }\href@noop {} {\bibfield  {journal} {\bibinfo  {journal} {Phys.
  Rev. D}\ }\textbf {\bibinfo {volume} {98}},\ \bibinfo {pages} {125006}
  (\bibinfo {year} {2018})}\BibitemShut {NoStop}%
\bibitem [{\citenamefont {Litim}\ \emph {et~al.}(2017)\citenamefont {Litim},
  \citenamefont {Marchais},\ and\ \citenamefont {Mati}}]{litim2017fixed}%
  \BibitemOpen
  \bibfield  {author} {\bibinfo {author} {\bibfnamefont {D.~F.}\ \bibnamefont
  {Litim}}, \bibinfo {author} {\bibfnamefont {E.}~\bibnamefont {Marchais}}, \
  and\ \bibinfo {author} {\bibfnamefont {P.}~\bibnamefont {Mati}},\ }\href@noop
  {} {\bibfield  {journal} {\bibinfo  {journal} {Phys. Rev. D}\ }\textbf
  {\bibinfo {volume} {95}},\ \bibinfo {pages} {125006} (\bibinfo {year}
  {2017})}\BibitemShut {NoStop}%
\bibitem [{\citenamefont {Amit}\ and\ \citenamefont
  {Rabinovici}(1985)}]{AMIT1985371}%
  \BibitemOpen
  \bibfield  {author} {\bibinfo {author} {\bibfnamefont {D.~J.}\ \bibnamefont
  {Amit}}\ and\ \bibinfo {author} {\bibfnamefont {E.}~\bibnamefont
  {Rabinovici}},\ }\href {\doibase
  https://doi.org/10.1016/0550-3213(85)90351-7} {\bibfield  {journal} {\bibinfo
   {journal} {Nucl. Phys. B}\ }\textbf {\bibinfo {volume} {257}},\ \bibinfo
  {pages} {371 } (\bibinfo {year} {1985})}\BibitemShut {NoStop}%
\bibitem [{\citenamefont {Bardeen}\ \emph {et~al.}(1985)\citenamefont
  {Bardeen}, \citenamefont {Higashijima},\ and\ \citenamefont
  {Moshe}}]{bardeen1985spontaneous}%
  \BibitemOpen
  \bibfield  {author} {\bibinfo {author} {\bibfnamefont {W.~A.}\ \bibnamefont
  {Bardeen}}, \bibinfo {author} {\bibfnamefont {K.}~\bibnamefont
  {Higashijima}}, \ and\ \bibinfo {author} {\bibfnamefont {M.}~\bibnamefont
  {Moshe}},\ }\href@noop {} {\bibfield  {journal} {\bibinfo  {journal} {Nucl.
  Phys. B}\ }\textbf {\bibinfo {volume} {250}},\ \bibinfo {pages} {437}
  (\bibinfo {year} {1985})}\BibitemShut {NoStop}%
\bibitem [{\citenamefont {Gudmundsdottir}\ and\ \citenamefont
  {Rydnell}(1985)}]{gudmundsdottir1985supersymmetric}%
  \BibitemOpen
  \bibfield  {author} {\bibinfo {author} {\bibfnamefont {R.}~\bibnamefont
  {Gudmundsdottir}}\ and\ \bibinfo {author} {\bibfnamefont {G.}~\bibnamefont
  {Rydnell}},\ }\href@noop {} {\bibfield  {journal} {\bibinfo  {journal} {Nucl.
  Phys. B}\ }\textbf {\bibinfo {volume} {254}},\ \bibinfo {pages} {593}
  (\bibinfo {year} {1985})}\BibitemShut {NoStop}%
\bibitem [{\citenamefont {Suzuki}\ and\ \citenamefont
  {Yamamoto}(1986)}]{suzuki1986nontrivial}%
  \BibitemOpen
  \bibfield  {author} {\bibinfo {author} {\bibfnamefont {T.}~\bibnamefont
  {Suzuki}}\ and\ \bibinfo {author} {\bibfnamefont {H.}~\bibnamefont
  {Yamamoto}},\ }\href@noop {} {\bibfield  {journal} {\bibinfo  {journal}
  {Prog. Theor. Phys.}\ }\textbf {\bibinfo {volume} {75}},\ \bibinfo {pages}
  {126} (\bibinfo {year} {1986})}\BibitemShut {NoStop}%
\bibitem [{\citenamefont {Suzuki}(1985)}]{suzuki1985three}%
  \BibitemOpen
  \bibfield  {author} {\bibinfo {author} {\bibfnamefont {T.}~\bibnamefont
  {Suzuki}},\ }\href@noop {} {\bibfield  {journal} {\bibinfo  {journal} {Phys.
  Rev. D}\ }\textbf {\bibinfo {volume} {32}},\ \bibinfo {pages} {1017}
  (\bibinfo {year} {1985})}\BibitemShut {NoStop}%
\bibitem [{\citenamefont {Gudmundsdottir}\ \emph {et~al.}(1984)\citenamefont
  {Gudmundsdottir}, \citenamefont {Rydnell},\ and\ \citenamefont
  {Salomonson}}]{gudmundsdottir1984more}%
  \BibitemOpen
  \bibfield  {author} {\bibinfo {author} {\bibfnamefont {R.}~\bibnamefont
  {Gudmundsdottir}}, \bibinfo {author} {\bibfnamefont {G.}~\bibnamefont
  {Rydnell}}, \ and\ \bibinfo {author} {\bibfnamefont {P.}~\bibnamefont
  {Salomonson}},\ }\href {\doibase 10.1103/PhysRevLett.53.2529} {\bibfield
  {journal} {\bibinfo  {journal} {Phys. Rev. Lett.}\ }\textbf {\bibinfo
  {volume} {53}},\ \bibinfo {pages} {2529} (\bibinfo {year}
  {1984})}\BibitemShut {NoStop}%
\bibitem [{\citenamefont {Matsubara}\ \emph {et~al.}(1985)\citenamefont
  {Matsubara}, \citenamefont {Suzuki},\ and\ \citenamefont
  {Yotsuyanagi}}]{matsubara1985large}%
  \BibitemOpen
  \bibfield  {author} {\bibinfo {author} {\bibfnamefont {Y.}~\bibnamefont
  {Matsubara}}, \bibinfo {author} {\bibfnamefont {T.}~\bibnamefont {Suzuki}}, \
  and\ \bibinfo {author} {\bibfnamefont {I.}~\bibnamefont {Yotsuyanagi}},\
  }\href@noop {} {\bibfield  {journal} {\bibinfo  {journal} {Z. Phys. C}\
  }\textbf {\bibinfo {volume} {27}},\ \bibinfo {pages} {599} (\bibinfo {year}
  {1985})}\BibitemShut {NoStop}%
\bibitem [{\citenamefont {Kessler}\ and\ \citenamefont
  {Neuberger}(1985)}]{kessler1985infinite}%
  \BibitemOpen
  \bibfield  {author} {\bibinfo {author} {\bibfnamefont {D.~A.}\ \bibnamefont
  {Kessler}}\ and\ \bibinfo {author} {\bibfnamefont {H.}~\bibnamefont
  {Neuberger}},\ }\href {\doibase https://doi.org/10.1016/0370-2693(85)90392-2}
  {\bibfield  {journal} {\bibinfo  {journal} {Phys. Lett. B}\ }\textbf
  {\bibinfo {volume} {157}},\ \bibinfo {pages} {416 } (\bibinfo {year}
  {1985})}\BibitemShut {NoStop}%
\bibitem [{\citenamefont {Stephen}\ and\ \citenamefont
  {McCauley}(1973)}]{STEPHEN197389}%
  \BibitemOpen
  \bibfield  {author} {\bibinfo {author} {\bibfnamefont {M.}~\bibnamefont
  {Stephen}}\ and\ \bibinfo {author} {\bibfnamefont {J.}~\bibnamefont
  {McCauley}},\ }\href {\doibase https://doi.org/10.1016/0375-9601(73)90799-8}
  {\bibfield  {journal} {\bibinfo  {journal} {Phys. Lett. A}\ }\textbf
  {\bibinfo {volume} {44}},\ \bibinfo {pages} {89 } (\bibinfo {year}
  {1973})}\BibitemShut {NoStop}%
\bibitem [{\citenamefont {Lewis}\ and\ \citenamefont
  {Adams}(1978)}]{lewis1978tricritical}%
  \BibitemOpen
  \bibfield  {author} {\bibinfo {author} {\bibfnamefont {A.}~\bibnamefont
  {Lewis}}\ and\ \bibinfo {author} {\bibfnamefont {F.}~\bibnamefont {Adams}},\
  }\href@noop {} {\bibfield  {journal} {\bibinfo  {journal} {Phys. Rev. B}\
  }\textbf {\bibinfo {volume} {18}},\ \bibinfo {pages} {5099} (\bibinfo {year}
  {1978})}\BibitemShut {NoStop}%
\bibitem [{\citenamefont {Pisarski}(1982)}]{pisarski1982fixed}%
  \BibitemOpen
  \bibfield  {author} {\bibinfo {author} {\bibfnamefont {R.~D.}\ \bibnamefont
  {Pisarski}},\ }\href@noop {} {\bibfield  {journal} {\bibinfo  {journal}
  {Phys. Rev. Lett.}\ }\textbf {\bibinfo {volume} {48}},\ \bibinfo {pages}
  {574} (\bibinfo {year} {1982})}\BibitemShut {NoStop}%
\bibitem [{\citenamefont {Osborn}\ and\ \citenamefont
  {Stergiou}(2018)}]{osborn2018seeking}%
  \BibitemOpen
  \bibfield  {author} {\bibinfo {author} {\bibfnamefont {H.}~\bibnamefont
  {Osborn}}\ and\ \bibinfo {author} {\bibfnamefont {A.}~\bibnamefont
  {Stergiou}},\ }\href@noop {} {\bibfield  {journal} {\bibinfo  {journal} {J.
  High Energy Phys.}\ }\textbf {\bibinfo {volume} {2018}},\ \bibinfo {pages}
  {51} (\bibinfo {year} {2018})}\BibitemShut {NoStop}%
\bibitem [{\citenamefont {Hager}(2002)}]{Hager_2002}%
  \BibitemOpen
  \bibfield  {author} {\bibinfo {author} {\bibfnamefont {J.~S.}\ \bibnamefont
  {Hager}},\ }\href {\doibase 10.1088/0305-4470/35/12/301} {\bibfield
  {journal} {\bibinfo  {journal} {J. Phys. A}\ }\textbf {\bibinfo {volume}
  {35}},\ \bibinfo {pages} {2703} (\bibinfo {year} {2002})}\BibitemShut
  {NoStop}%
\bibitem [{\citenamefont {Pisarski}(1983)}]{Pisarski2}%
  \BibitemOpen
  \bibfield  {author} {\bibinfo {author} {\bibfnamefont {R.~D.}\ \bibnamefont
  {Pisarski}},\ }\href {\doibase 10.1103/PhysRevD.28.1554} {\bibfield
  {journal} {\bibinfo  {journal} {Phys. Rev. D}\ }\textbf {\bibinfo {volume}
  {28}},\ \bibinfo {pages} {1554} (\bibinfo {year} {1983})}\BibitemShut
  {NoStop}%
\bibitem [{\citenamefont {Yabunaka}\ and\ \citenamefont
  {Delamotte}(2017)}]{yabunaka2017surprises}%
  \BibitemOpen
  \bibfield  {author} {\bibinfo {author} {\bibfnamefont {S.}~\bibnamefont
  {Yabunaka}}\ and\ \bibinfo {author} {\bibfnamefont {B.}~\bibnamefont
  {Delamotte}},\ }\href@noop {} {\bibfield  {journal} {\bibinfo  {journal}
  {Phys. Rev. Lett.}\ }\textbf {\bibinfo {volume} {119}},\ \bibinfo {pages}
  {191602} (\bibinfo {year} {2017})}\BibitemShut {NoStop}%
\bibitem [{\citenamefont {Yabunaka}\ and\ \citenamefont
  {Delamotte}(2018)}]{yabunaka2018might}%
  \BibitemOpen
  \bibfield  {author} {\bibinfo {author} {\bibfnamefont {S.}~\bibnamefont
  {Yabunaka}}\ and\ \bibinfo {author} {\bibfnamefont {B.}~\bibnamefont
  {Delamotte}},\ }\href@noop {} {\bibfield  {journal} {\bibinfo  {journal}
  {Phys. Rev. Lett.}\ }\textbf {\bibinfo {volume} {121}},\ \bibinfo {pages}
  {231601} (\bibinfo {year} {2018})}\BibitemShut {NoStop}%
\bibitem [{\citenamefont {Tissier}\ and\ \citenamefont
  {Tarjus}(2008)}]{tissier08}%
  \BibitemOpen
  \bibfield  {author} {\bibinfo {author} {\bibfnamefont {M.}~\bibnamefont
  {Tissier}}\ and\ \bibinfo {author} {\bibfnamefont {G.}~\bibnamefont
  {Tarjus}},\ }\href@noop {} {\bibfield  {journal} {\bibinfo  {journal} {Phys.
  Rev. B}\ }\textbf {\bibinfo {volume} {78}},\ \bibinfo {pages} {024204}
  (\bibinfo {year} {2008})}\BibitemShut {NoStop}%
\bibitem [{\citenamefont {Tissier}\ and\ \citenamefont
  {Wschebor}(2010)}]{tissier10}%
  \BibitemOpen
  \bibfield  {author} {\bibinfo {author} {\bibfnamefont {M.}~\bibnamefont
  {Tissier}}\ and\ \bibinfo {author} {\bibfnamefont {N.}~\bibnamefont
  {Wschebor}},\ }\href@noop {} {\bibfield  {journal} {\bibinfo  {journal}
  {Phys. Rev. D}\ }\textbf {\bibinfo {volume} {82}},\ \bibinfo {pages} {101701}
  (\bibinfo {year} {2010})}\BibitemShut {NoStop}%
\bibitem [{\citenamefont {Tissier}\ and\ \citenamefont
  {Tarjus}(2006)}]{TissierPhysRevB.74.214419}%
  \BibitemOpen
  \bibfield  {author} {\bibinfo {author} {\bibfnamefont {M.}~\bibnamefont
  {Tissier}}\ and\ \bibinfo {author} {\bibfnamefont {G.}~\bibnamefont
  {Tarjus}},\ }\href {\doibase 10.1103/PhysRevB.74.214419} {\bibfield
  {journal} {\bibinfo  {journal} {Phys. Rev. B}\ }\textbf {\bibinfo {volume}
  {74}},\ \bibinfo {pages} {214419} (\bibinfo {year} {2006})}\BibitemShut
  {NoStop}%
\bibitem [{\citenamefont {Canet}\ \emph
  {et~al.}(2016{\natexlab{a}})\citenamefont {Canet}, \citenamefont
  {Delamotte},\ and\ \citenamefont {Wschebor}}]{CanetPhysRevE.93.063101}%
  \BibitemOpen
  \bibfield  {author} {\bibinfo {author} {\bibfnamefont {L.}~\bibnamefont
  {Canet}}, \bibinfo {author} {\bibfnamefont {B.}~\bibnamefont {Delamotte}}, \
  and\ \bibinfo {author} {\bibfnamefont {N.}~\bibnamefont {Wschebor}},\ }\href
  {\doibase 10.1103/PhysRevE.93.063101} {\bibfield  {journal} {\bibinfo
  {journal} {Phys. Rev. E}\ }\textbf {\bibinfo {volume} {93}},\ \bibinfo
  {pages} {063101} (\bibinfo {year} {2016}{\natexlab{a}})}\BibitemShut
  {NoStop}%
\bibitem [{\citenamefont {Peles}\ \emph {et~al.}(2004)\citenamefont {Peles},
  \citenamefont {Southern}, \citenamefont {Delamotte}, \citenamefont
  {Mouhanna},\ and\ \citenamefont {Tissier}}]{PhysRevB.69.220408}%
  \BibitemOpen
  \bibfield  {author} {\bibinfo {author} {\bibfnamefont {A.}~\bibnamefont
  {Peles}}, \bibinfo {author} {\bibfnamefont {B.~W.}\ \bibnamefont {Southern}},
  \bibinfo {author} {\bibfnamefont {B.}~\bibnamefont {Delamotte}}, \bibinfo
  {author} {\bibfnamefont {D.}~\bibnamefont {Mouhanna}}, \ and\ \bibinfo
  {author} {\bibfnamefont {M.}~\bibnamefont {Tissier}},\ }\href {\doibase
  10.1103/PhysRevB.69.220408} {\bibfield  {journal} {\bibinfo  {journal} {Phys.
  Rev. B}\ }\textbf {\bibinfo {volume} {69}},\ \bibinfo {pages} {220408}
  (\bibinfo {year} {2004})}\BibitemShut {NoStop}%
\bibitem [{\citenamefont {Caffarel}\ \emph {et~al.}(2001)\citenamefont
  {Caffarel}, \citenamefont {Azaria}, \citenamefont {Delamotte},\ and\
  \citenamefont {Mouhanna}}]{PhysRevB.64.014412}%
  \BibitemOpen
  \bibfield  {author} {\bibinfo {author} {\bibfnamefont {M.}~\bibnamefont
  {Caffarel}}, \bibinfo {author} {\bibfnamefont {P.}~\bibnamefont {Azaria}},
  \bibinfo {author} {\bibfnamefont {B.}~\bibnamefont {Delamotte}}, \ and\
  \bibinfo {author} {\bibfnamefont {D.}~\bibnamefont {Mouhanna}},\ }\href
  {\doibase 10.1103/PhysRevB.64.014412} {\bibfield  {journal} {\bibinfo
  {journal} {Phys. Rev. B}\ }\textbf {\bibinfo {volume} {64}},\ \bibinfo
  {pages} {014412} (\bibinfo {year} {2001})}\BibitemShut {NoStop}%
\bibitem [{\citenamefont {Canet}\ \emph {et~al.}(2011)\citenamefont {Canet},
  \citenamefont {Chat\'e}, \citenamefont {Delamotte},\ and\ \citenamefont
  {Wschebor}}]{CanetPhysRevE.84.061128}%
  \BibitemOpen
  \bibfield  {author} {\bibinfo {author} {\bibfnamefont {L.}~\bibnamefont
  {Canet}}, \bibinfo {author} {\bibfnamefont {H.}~\bibnamefont {Chat\'e}},
  \bibinfo {author} {\bibfnamefont {B.}~\bibnamefont {Delamotte}}, \ and\
  \bibinfo {author} {\bibfnamefont {N.}~\bibnamefont {Wschebor}},\ }\href
  {\doibase 10.1103/PhysRevE.84.061128} {\bibfield  {journal} {\bibinfo
  {journal} {Phys. Rev. E}\ }\textbf {\bibinfo {volume} {84}},\ \bibinfo
  {pages} {061128} (\bibinfo {year} {2011})}\BibitemShut {NoStop}%
\bibitem [{\citenamefont {Gredat}\ \emph {et~al.}(2014)\citenamefont {Gredat},
  \citenamefont {Chat\'e}, \citenamefont {Delamotte},\ and\ \citenamefont
  {Dornic}}]{GredatPhysRevE.89.010102}%
  \BibitemOpen
  \bibfield  {author} {\bibinfo {author} {\bibfnamefont {D.}~\bibnamefont
  {Gredat}}, \bibinfo {author} {\bibfnamefont {H.}~\bibnamefont {Chat\'e}},
  \bibinfo {author} {\bibfnamefont {B.}~\bibnamefont {Delamotte}}, \ and\
  \bibinfo {author} {\bibfnamefont {I.}~\bibnamefont {Dornic}},\ }\href
  {\doibase 10.1103/PhysRevE.89.010102} {\bibfield  {journal} {\bibinfo
  {journal} {Phys. Rev. E}\ }\textbf {\bibinfo {volume} {89}},\ \bibinfo
  {pages} {010102} (\bibinfo {year} {2014})}\BibitemShut {NoStop}%
\bibitem [{\citenamefont {Wilson}(1971)}]{PhysRevB.4.3174}%
  \BibitemOpen
  \bibfield  {author} {\bibinfo {author} {\bibfnamefont {K.}~\bibnamefont
  {Wilson}},\ }\href@noop {} {\bibfield  {journal} {\bibinfo  {journal} {Phys.
  Rev. B}\ }\textbf {\bibinfo {volume} {4}},\ \bibinfo {pages} {3174} (\bibinfo
  {year} {1971})}\BibitemShut {NoStop}%
\bibitem [{\citenamefont {Wetterich}(1991)}]{wetterich91}%
  \BibitemOpen
  \bibfield  {author} {\bibinfo {author} {\bibfnamefont {C.}~\bibnamefont
  {Wetterich}},\ }\href@noop {} {\bibfield  {journal} {\bibinfo  {journal}
  {Nucl. Phys. B}\ }\textbf {\bibinfo {volume} {352}},\ \bibinfo {pages} {529}
  (\bibinfo {year} {1991})}\BibitemShut {NoStop}%
\bibitem [{\citenamefont {Wetterich}(1993)}]{wetterich93b}%
  \BibitemOpen
  \bibfield  {author} {\bibinfo {author} {\bibfnamefont {C.}~\bibnamefont
  {Wetterich}},\ }\href@noop {} {\bibfield  {journal} {\bibinfo  {journal}
  {Phys. Lett. B}\ }\textbf {\bibinfo {volume} {301}},\ \bibinfo {pages} {90}
  (\bibinfo {year} {1993})}\BibitemShut {NoStop}%
\bibitem [{\citenamefont {Ellwanger}(1993)}]{Ellwanger}%
  \BibitemOpen
  \bibfield  {author} {\bibinfo {author} {\bibfnamefont {U.}~\bibnamefont
  {Ellwanger}},\ }\href@noop {} {\bibfield  {journal} {\bibinfo  {journal} {Z.
  Phys. C}\ }\textbf {\bibinfo {volume} {58}},\ \bibinfo {pages} {619}
  (\bibinfo {year} {1993})}\BibitemShut {NoStop}%
\bibitem [{\citenamefont {Morris}(1994)}]{Morris94}%
  \BibitemOpen
  \bibfield  {author} {\bibinfo {author} {\bibfnamefont {T.~R.}\ \bibnamefont
  {Morris}},\ }\href@noop {} {\bibfield  {journal} {\bibinfo  {journal} {Z.
  Phys. C}\ }\textbf {\bibinfo {volume} {9}},\ \bibinfo {pages} {2411}
  (\bibinfo {year} {1994})}\BibitemShut {NoStop}%
\bibitem [{\citenamefont {Delamotte}(2012)}]{delamotte2012}%
  \BibitemOpen
  \bibfield  {author} {\bibinfo {author} {\bibfnamefont {B.}~\bibnamefont
  {Delamotte}},\ }\href@noop {} {\bibfield  {journal} {\bibinfo  {journal}
  {Lect. Notes Phys.}\ }\textbf {\bibinfo {volume} {852}},\ \bibinfo {pages}
  {49} (\bibinfo {year} {2012})}\BibitemShut {NoStop}%
\bibitem [{\citenamefont {Berges}\ \emph {et~al.}(2002)\citenamefont {Berges},
  \citenamefont {Tetradis},\ and\ \citenamefont {Wetterich}}]{berges2002}%
  \BibitemOpen
  \bibfield  {author} {\bibinfo {author} {\bibfnamefont {J.}~\bibnamefont
  {Berges}}, \bibinfo {author} {\bibfnamefont {N.}~\bibnamefont {Tetradis}}, \
  and\ \bibinfo {author} {\bibfnamefont {C.}~\bibnamefont {Wetterich}},\
  }\href@noop {} {\bibfield  {journal} {\bibinfo  {journal} {Phys. Rep.}\
  }\textbf {\bibinfo {volume} {363}},\ \bibinfo {pages} {223} (\bibinfo {year}
  {2002})}\BibitemShut {NoStop}%
\bibitem [{\citenamefont {Delamotte}\ and\ \citenamefont
  {Canet}(2005)}]{delamotte04}%
  \BibitemOpen
  \bibfield  {author} {\bibinfo {author} {\bibfnamefont {B.}~\bibnamefont
  {Delamotte}}\ and\ \bibinfo {author} {\bibfnamefont {L.}~\bibnamefont
  {Canet}},\ }\href@noop {} {\bibfield  {journal} {\bibinfo  {journal}
  {Condens. Matter Phys.}\ }\textbf {\bibinfo {volume} {8}},\ \bibinfo {pages}
  {163} (\bibinfo {year} {2005})}\BibitemShut {NoStop}%
\bibitem [{\citenamefont {Canet}\ \emph {et~al.}(2004)\citenamefont {Canet},
  \citenamefont {Delamotte}, \citenamefont {Deloubri{\`e}re},\ and\
  \citenamefont {Wschebor}}]{canet04}%
  \BibitemOpen
  \bibfield  {author} {\bibinfo {author} {\bibfnamefont {L.}~\bibnamefont
  {Canet}}, \bibinfo {author} {\bibfnamefont {B.}~\bibnamefont {Delamotte}},
  \bibinfo {author} {\bibfnamefont {O.}~\bibnamefont {Deloubri{\`e}re}}, \ and\
  \bibinfo {author} {\bibfnamefont {N.}~\bibnamefont {Wschebor}},\ }\href@noop
  {} {\bibfield  {journal} {\bibinfo  {journal} {Phys. Rev. Lett.}\ }\textbf
  {\bibinfo {volume} {92}},\ \bibinfo {pages} {195703} (\bibinfo {year}
  {2004})}\BibitemShut {NoStop}%
\bibitem [{\citenamefont {Kloss}\ \emph {et~al.}(2014)\citenamefont {Kloss},
  \citenamefont {Canet}, \citenamefont {Delamotte},\ and\ \citenamefont
  {Wschebor}}]{kloss14}%
  \BibitemOpen
  \bibfield  {author} {\bibinfo {author} {\bibfnamefont {T.}~\bibnamefont
  {Kloss}}, \bibinfo {author} {\bibfnamefont {L.}~\bibnamefont {Canet}},
  \bibinfo {author} {\bibfnamefont {B.}~\bibnamefont {Delamotte}}, \ and\
  \bibinfo {author} {\bibfnamefont {N.}~\bibnamefont {Wschebor}},\ }\href@noop
  {} {\bibfield  {journal} {\bibinfo  {journal} {Phys. Rev. E}\ }\textbf
  {\bibinfo {volume} {89}},\ \bibinfo {pages} {022108} (\bibinfo {year}
  {2014})}\BibitemShut {NoStop}%
\bibitem [{\citenamefont {Canet}\ \emph
  {et~al.}(2016{\natexlab{b}})\citenamefont {Canet}, \citenamefont
  {Delamotte},\ and\ \citenamefont {Wschebor}}]{canet16}%
  \BibitemOpen
  \bibfield  {author} {\bibinfo {author} {\bibfnamefont {L.}~\bibnamefont
  {Canet}}, \bibinfo {author} {\bibfnamefont {B.}~\bibnamefont {Delamotte}}, \
  and\ \bibinfo {author} {\bibfnamefont {N.}~\bibnamefont {Wschebor}},\ }\href
  {\doibase 10.1103/PhysRevE.93.063101} {\bibfield  {journal} {\bibinfo
  {journal} {Phys. Rev. E}\ }\textbf {\bibinfo {volume} {93}},\ \bibinfo
  {pages} {063101} (\bibinfo {year} {2016}{\natexlab{b}})}\BibitemShut
  {NoStop}%
\bibitem [{\citenamefont {Morris}(2005)}]{Morris}%
  \BibitemOpen
  \bibfield  {author} {\bibinfo {author} {\bibfnamefont {T.~R.}\ \bibnamefont
  {Morris}},\ }\href@noop {} {\bibfield  {journal} {\bibinfo  {journal} {JHEP}\
  }\textbf {\bibinfo {volume} {2005}},\ \bibinfo {pages} {027} (\bibinfo {year}
  {2005})}\BibitemShut {NoStop}%
\bibitem [{\citenamefont {Omid}\ \emph {et~al.}(2016)\citenamefont {Omid},
  \citenamefont {Semenoff},\ and\ \citenamefont {Wijewardhana}}]{Omid}%
  \BibitemOpen
  \bibfield  {author} {\bibinfo {author} {\bibfnamefont {H.}~\bibnamefont
  {Omid}}, \bibinfo {author} {\bibfnamefont {G.~W.}\ \bibnamefont {Semenoff}},
  \ and\ \bibinfo {author} {\bibfnamefont {L.}~\bibnamefont {Wijewardhana}},\
  }\href@noop {} {\bibfield  {journal} {\bibinfo  {journal} {Phys. Rev. D}\
  }\textbf {\bibinfo {volume} {94}},\ \bibinfo {pages} {125017} (\bibinfo
  {year} {2016})}\BibitemShut {NoStop}%
\bibitem [{Note1()}]{Note1}%
  \BibitemOpen
  \bibinfo {note} {Notice that the linear part of the BMB FP potential
  corresponding to $\protect \mathaccentV {bar}016\varrho \in [0,\protect
  \mathaccentV {bar}016\varrho _0]$, see Fig. \ref {V(rho)}, can be replaced by
  a smooth analytic continuation of the potential obtained for $\protect
  \mathaccentV {bar}016\varrho >\protect \mathaccentV {bar}016\varrho _0$.
  Considering either solution has no physical consequence because in both
  cases, the interval $\protect \mathaccentV {bar}016\varrho \in [\protect
  \mathaccentV {bar}016\varrho _0,\infty [$ is entirely mapped onto $\protect
  \mathaccentV {bar}016\rho \ge 0$ in the Wetterich version of the
  flow.}\BibitemShut {Stop}%
\bibitem [{Note2()}]{Note2}%
  \BibitemOpen
  \bibinfo {note} {In \cite {yabunaka2018might}, $SA_3$ was called
  $C_3$.}\BibitemShut {Stop}%
\bibitem [{\citenamefont {Canet}\ \emph {et~al.}(2005)\citenamefont {Canet},
  \citenamefont {Chat\'e}, \citenamefont {Delamotte}, \citenamefont {Dornic},\
  and\ \citenamefont {Mu\~noz}}]{CanetPhysRevLett.95.100601}%
  \BibitemOpen
  \bibfield  {author} {\bibinfo {author} {\bibfnamefont {L.}~\bibnamefont
  {Canet}}, \bibinfo {author} {\bibfnamefont {H.}~\bibnamefont {Chat\'e}},
  \bibinfo {author} {\bibfnamefont {B.}~\bibnamefont {Delamotte}}, \bibinfo
  {author} {\bibfnamefont {I.}~\bibnamefont {Dornic}}, \ and\ \bibinfo {author}
  {\bibfnamefont {M.~A.}\ \bibnamefont {Mu\~noz}},\ }\href {\doibase
  10.1103/PhysRevLett.95.100601} {\bibfield  {journal} {\bibinfo  {journal}
  {Phys. Rev. Lett.}\ }\textbf {\bibinfo {volume} {95}},\ \bibinfo {pages}
  {100601} (\bibinfo {year} {2005})}\BibitemShut {NoStop}%
\bibitem [{\citenamefont {Canet}\ \emph {et~al.}(2003)\citenamefont {Canet},
  \citenamefont {Delamotte}, \citenamefont {Mouhanna},\ and\ \citenamefont
  {Vidal}}]{canet03}%
  \BibitemOpen
  \bibfield  {author} {\bibinfo {author} {\bibfnamefont {L.}~\bibnamefont
  {Canet}}, \bibinfo {author} {\bibfnamefont {B.}~\bibnamefont {Delamotte}},
  \bibinfo {author} {\bibfnamefont {D.}~\bibnamefont {Mouhanna}}, \ and\
  \bibinfo {author} {\bibfnamefont {J.}~\bibnamefont {Vidal}},\ }\href@noop {}
  {\bibfield  {journal} {\bibinfo  {journal} {Phys. Rev. B}\ }\textbf {\bibinfo
  {volume} {68}},\ \bibinfo {pages} {064421} (\bibinfo {year}
  {2003})}\BibitemShut {NoStop}%
\bibitem [{\citenamefont {Canet}(2005)}]{canet05}%
  \BibitemOpen
  \bibfield  {author} {\bibinfo {author} {\bibfnamefont {L.}~\bibnamefont
  {Canet}},\ }\href@noop {} {\bibfield  {journal} {\bibinfo  {journal} {Phys.
  Rev. B}\ }\textbf {\bibinfo {volume} {71}},\ \bibinfo {pages} {012418}
  (\bibinfo {year} {2005})}\BibitemShut {NoStop}%
\bibitem [{\citenamefont {L{\'e}onard}\ and\ \citenamefont
  {Delamotte}(2015)}]{leonard15}%
  \BibitemOpen
  \bibfield  {author} {\bibinfo {author} {\bibfnamefont {F.}~\bibnamefont
  {L{\'e}onard}}\ and\ \bibinfo {author} {\bibfnamefont {B.}~\bibnamefont
  {Delamotte}},\ }\href@noop {} {\bibfield  {journal} {\bibinfo  {journal}
  {Phys. Rev. Lett.}\ }\textbf {\bibinfo {volume} {115}},\ \bibinfo {pages}
  {200601} (\bibinfo {year} {2015})}\BibitemShut {NoStop}%
\bibitem [{\citenamefont {Jakubczyk}\ \emph {et~al.}(2014)\citenamefont
  {Jakubczyk}, \citenamefont {Dupuis},\ and\ \citenamefont
  {Delamotte}}]{JakubczykPhysRevE.90.062105}%
  \BibitemOpen
  \bibfield  {author} {\bibinfo {author} {\bibfnamefont {P.}~\bibnamefont
  {Jakubczyk}}, \bibinfo {author} {\bibfnamefont {N.}~\bibnamefont {Dupuis}}, \
  and\ \bibinfo {author} {\bibfnamefont {B.}~\bibnamefont {Delamotte}},\ }\href
  {\doibase 10.1103/PhysRevE.90.062105} {\bibfield  {journal} {\bibinfo
  {journal} {Phys. Rev. E}\ }\textbf {\bibinfo {volume} {90}},\ \bibinfo
  {pages} {062105} (\bibinfo {year} {2014})}\BibitemShut {NoStop}%
\bibitem [{\citenamefont {Balog}\ \emph {et~al.}(2019)\citenamefont {Balog},
  \citenamefont {Chat\'e}, \citenamefont {Delamotte}, \citenamefont
  {Marohni\'c},\ and\ \citenamefont {Wschebor}}]{Balog2019}%
  \BibitemOpen
  \bibfield  {author} {\bibinfo {author} {\bibfnamefont {I.}~\bibnamefont
  {Balog}}, \bibinfo {author} {\bibfnamefont {H.}~\bibnamefont {Chat\'e}},
  \bibinfo {author} {\bibfnamefont {B.}~\bibnamefont {Delamotte}}, \bibinfo
  {author} {\bibfnamefont {M.}~\bibnamefont {Marohni\'c}}, \ and\ \bibinfo
  {author} {\bibfnamefont {N.}~\bibnamefont {Wschebor}},\ }\href
  {https://link.aps.org/doi.org/10.1103/PhysRevLett.123.240604} {\bibfield
  {journal} {\bibinfo  {journal} {Phys. Rev. Lett.}\ }\textbf {\bibinfo
  {volume} {123}},\ \bibinfo {pages} {240604} (\bibinfo {year}
  {2019})}\BibitemShut {NoStop}%
\bibitem [{\citenamefont {Fleming}()}]{fleming}%
  \BibitemOpen
  \bibfield  {author} {\bibinfo {author} {\bibfnamefont {C.}~\bibnamefont
  {Fleming}},\ }\href@noop {} {\bibinfo  {journal} {in preparation}\
  }\BibitemShut {NoStop}%
\bibitem [{Note3()}]{Note3}%
  \BibitemOpen
\bibfield  {journal} {  }\bibinfo {note} {One may also expand around the finite
  $N$ minimum $\protect \mathaccentV {bar}016{\varrho }_m$ of $v$. However,
  this point being a function of $\epsilon $, the expansion turns out to be in
  powers of $\epsilon ^{1/2}$ which makes the calculations a little more
  difficult.}\BibitemShut {Stop}%
\end{thebibliography}%

\pagebreak


\onecolumngrid

\begin{center}
\textbf{\large Supplemental Material}
\end{center}
\setcounter{equation}{0}
\setcounter{figure}{0}
\setcounter{table}{0}
\setcounter{page}{1}
\renewcommand{\theequation}{S\arabic{equation}}
\renewcommand{\thefigure}{S\arabic{figure}}
\renewcommand{\bibnumfmt}[1]{[S#1]}
\renewcommand{\citenumfont}[1]{S#1}

\section{Plot of $\bar V'$ and singularity of the BMB FP potential}

In Fig.1 of the main text, for $\tau=\tau_{\text{BMB}}$, $\bar{V}(\bar{\varrho})$ has a discontinuous curvature that is not very visible. We thus choose to plot $\bar{V}'$ as a function of $\bar{\varrho}$:

\begin{figure}[h]
\begin{centering}
\includegraphics[width=0.5\linewidth]{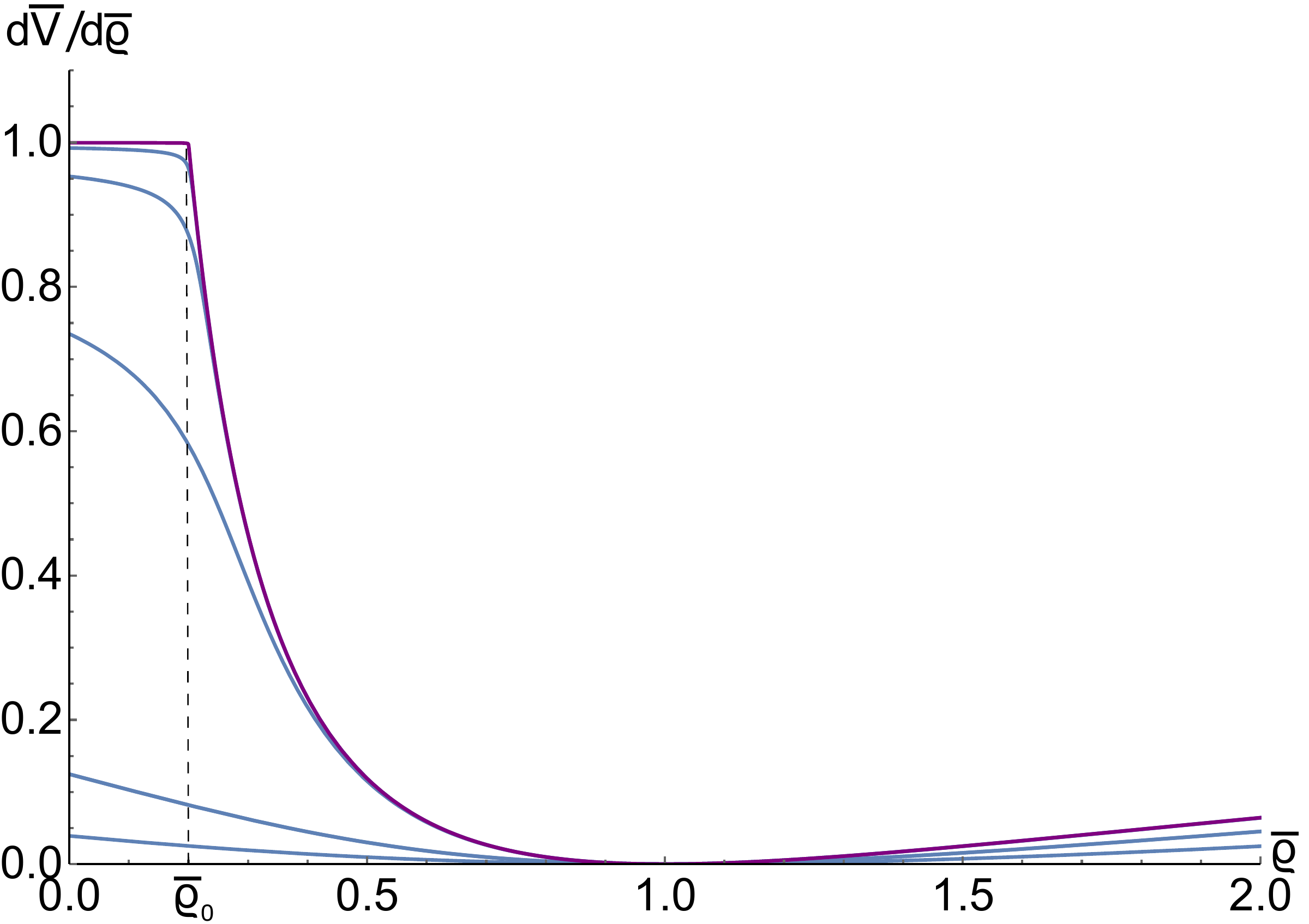} 
\par\end{centering}
\caption{Potentials $\bar V'(\bar\varrho)$ of the tricritical FPs ${\cal A}(\tau)=\{A(\tau),\tilde A(\tau)\}$ of the BMB line (blue). The BMB FP is the endpoint of the BMB line (purple). The second derivative $\bar V''(\bar\varrho)$ of the potential of the BMB FP shows a discontinuity in its second derivative at $\bar\varrho=\bar\varrho_0$.
}
\label{figVp}
\end{figure}

\section{Plot of $\tau$ as a function of $\alpha$}

We plot here the relation between $\tau$ and $\alpha$ as a visual means of understanding how the finite $N$ FPs $A_2(\alpha)$ and $\tilde{A}_3(\alpha)$ relate to the BMB line ${\cal A}(\tau)$

\begin{figure}[h]
\begin{centering}
\includegraphics[scale=0.35]{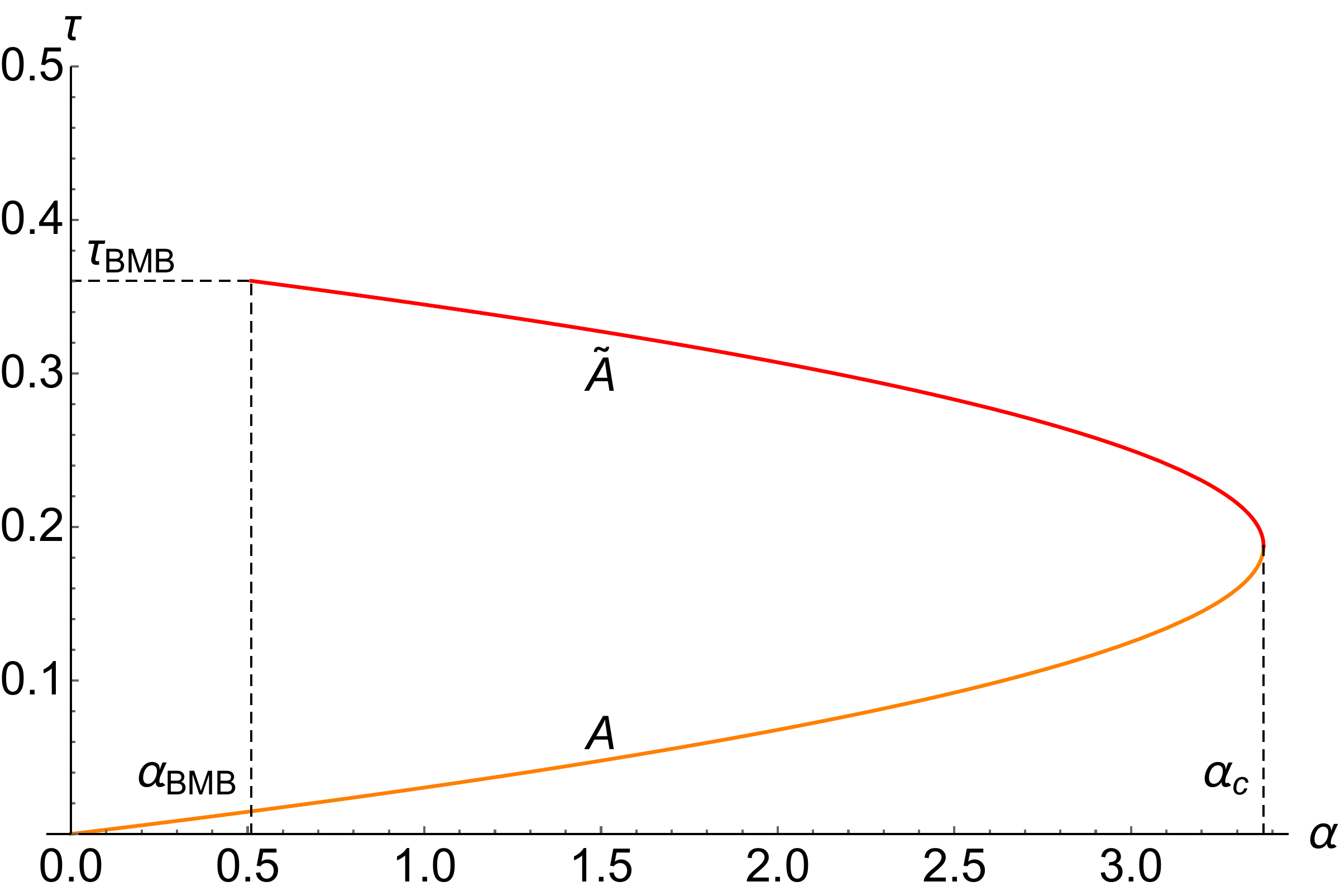} 
\par\end{centering}
\caption{$\tau$ on the BMB line at $N=\infty$ as a function of $\alpha=(d-3)N$ within the LPA approximation of the NPRG formalism. The two branches given by $A$ and $\tilde A$ give the limits when $N\rightarrow\infty$ of the two FPs at finite but large $N$, namely $A_2(\alpha)$ and $\tilde{A}_3(\alpha)$. Both branches meet at $\alpha=\alpha_{c}=3.375$ and the upper branch $\tilde A$ extends to the point $(\alpha_{\text{BMB}},\tau_{\text{BMB}})\simeq(0.51,0.36)$
}
\label{taualpha}
\end{figure}

\section{ Derivation of the relation between $\alpha$ and $\tau$, Eq. (9) of the main text}

We recall the LPA differential equation in the Wilson-Polchinski formulation $\rho$:
\begin{equation}
1-d\bar{V}+\left(d-2\right)\,\bar{\varrho}\,\bar{V}'+2\,\bar{\varrho}\,\bar{V}'^{2}-\,\bar{V}'-\frac{2}{N}\,\bar{\varrho}\,\bar{V}''=0.\label{eq:V}
\end{equation}

Deriving this equation with respect to $\bar{\varrho}$ and writing $v=\bar{V}'$
and $1/N=\epsilon$ we obtain:
\begin{equation}
((d-2)\bar{\varrho}-2\epsilon-1)v'-2\bar{\varrho}\epsilon v''+2v\left(2\bar{\varrho} v'+v-1\right)=0.\label{eq:veq}
\end{equation}

We now present two methods to obtain the relation  between $\alpha$ and $\tau$ given in Eq. (9) of the main text. The first method is straightforward and requires expanding the potential in powers of $\epsilon$ and $\bar{\varrho}-1$ \footnote{One may also expand around the finite $N$ minimum $\bar{\varrho}_m$ of $v$. However, this point being a function of $\epsilon$, the expansion turns out to be in powers of $\epsilon^{1/2}$ which makes the calculations a little more difficult.}. However, this method gives information only in the vicinity of $\bar{\varrho}=1$ and does not explain the type of non analytical behaviour obtained when one refuses to choose the relation between $\alpha$ and $\tau$. Thus, we will also use a fully functional method. This method has the advantage of yielding the potential at finite and large $N$ up to $1/N^2$ corrections and is therefore useful to get the behavior near $\bar{\varrho}=0$ where a divergence appears at $N=\infty$ and $\tau=\tau_{BMB}$.

The first method consists in Taylor expanding Eq.~(\ref{eq:veq}) about $\bar{\varrho}=1$ (the inflexion point)  with $v=\sum a_k (\bar{\varrho}-1)^k$. We moreover expand the couplings $a_k$ in powers of $\epsilon$ as $a_k=a_k^0+\epsilon a_k^1+O(\epsilon^2)$ where the $a_k^0$'s are the couplings involved in the expansion of the FP potential given at $N=\infty$ by Eq.~(9). The system of equations obtained by independently setting equal to 0  the coefficients of $\epsilon^n(\bar{\varrho}-1)^p$ yields the relation between $\alpha$ and $\tau$ given in Eq.~(9). Notice that in this method the $\tau$ dependence comes from the $a_k^0$'s.

The functional method consists in expanding $v$ as $v=v_{0}+\epsilon v_{1}+O(\epsilon^2)$ in Eq.(\ref{eq:veq}). At order $\epsilon$, this yields a differential equation on $v_1$ that depends on $\bar{\varrho}$, $v_0$, $v_0'$ and $v_0''$. Using Eq.(\ref{eq:veq}) and its derivative both evaluated at $\epsilon=0$, $v_0''$ and $v_0'$ can be eliminated in terms of $v_0$. This leads to:

\begin{equation}
\begin{aligned}32\bar{\varrho}^{2} v_0^{4}\left(\alpha\bar{\varrho}+4\bar{\varrho}\left(2\bar{\varrho}  v_1'+  v_1\right)-1\right)+16\bar{\varrho}^{2} v_0^{3}\left(-\alpha(\bar{\varrho}+1)+8(\bar{\varrho}-1)\left(2\bar{\varrho}  v_1'+  v_1\right)+1\right)+\\
2 v_0^{2}\left(\bar{\varrho}\left(-7\alpha\bar{\varrho}^{2}+6(\alpha+2)\bar{\varrho}+\alpha-6\right)+4(\bar{\varrho}-1)\bar{\varrho}\left(12(\bar{\varrho}-1)\bar{\varrho}  v_1'+(\bar{\varrho}-5)  v_1\right)+2\right)\\
2(\bar{\varrho}-1) v_0\left(\bar{\varrho}(\alpha(-\bar{\varrho})+\alpha-4)+8\bar{\varrho}(\bar{\varrho}-1)^{2}  v_1'+\left(-6\bar{\varrho}^{2}+4\bar{\varrho}+2\right)  v_1+2\right)+(\bar{\varrho}-1)^{3}\left((\bar{\varrho}-1)  v_1'-2  v_1\right)=0.
\end{aligned}\label{v1dif}
\end{equation}
We then assume that $v_1$ is analytic at $h=\bar{\varrho}-1=0$. The Taylor expansion of $v_0$
at $h=0$ is:

\begin{equation}
\begin{aligned}v_0= \frac{h^2 \tau }{2}-2 h^3 \tau ^2+\frac{5}{4} h^4 \tau ^2 (8 \tau -1)+h^5 (13-56 \tau ) \tau ^3+\\
\frac{7}{8} h^6 \tau ^3 (128 \tau  (3 \tau -1)+5)+h^7 \left(-2112 \tau ^6+912 \tau ^5-\frac{383 \tau ^4}{5}\right)+O\left(h^8\right).
\end{aligned}
\label{v0}
\end{equation}

Inserting Eq.(\ref{v0}) into Eq.(\ref{v1dif}), neglecting terms of order 5,  and dividing by $h^{3}$ 
gives :

\begin{equation}
-\tau h^2(\alpha +6 \tau  (12 \tau -5))+\left(8 h^2 \tau +h\right) \text{v1}'(x)+\left(6 h^2 \tau  (4 \tau -1)-8 h \tau -2\right) \text{v1}(x)-h \tau  (\alpha -12 \tau +4)-2 \tau=0.
\label{eqv1}
\end{equation}

Finally, replacing $v_1$  in Eq. (\ref{eqv1}) by its Taylor expansion: $v_1=v_1^{(0)}+v_1^{(1)}\,h+v_1^{(2)}\,h^{2}+O(h)^{3}$  leads to:

\begin{subequations}
\begin{align}v_1^{(0)} &=-\tau \label{systa}\\
v_1^{(1)} &=\tau  (-\alpha +20 \tau -4)\label{systb}\\
\alpha &= 36 \tau -96 \tau ^2.\label{systc}
\end{align}
\end{subequations}
Notice that it is because the $v_1^{(2)}$ term cancels in Eq. (\ref{systc}) that we can obtain a relation between $\alpha$ and $\tau$ only.

Let us finally notice that Eq.(\ref{systc}) can be retrieved in a functional way. The solution of Eq.(\ref{v1dif}) is:
\begin{equation}
\begin{aligned}v_1\left(\bar{\varrho}\right)=\exp\left(K\left(\bar{\varrho}\right)\right)\,\left(C-\int_{1}^{\bar{\varrho}}2e^{-K(\chi)}\left(\frac{16\alpha\chi^{3} v_0(\chi)^{4}-8\alpha\chi^{3} v_0(\chi)^{3}-7\alpha\chi^{3} v_0(\chi)^{2}-\alpha\chi^{3} v_0(\chi)-8\alpha\chi^{2} v_0(\chi)^{3}}{(4\chi v_0(\chi)+\chi-1)^{4}}\right.+\right.\\
\frac{6\alpha\chi^{2} v_0(\chi)^{2}+2\alpha\chi^{2} v_0(\chi)+\alpha\chi v_0(\chi)^{2}-\alpha\chi v_0(\chi)-16\chi^{2} v_0(\chi)^{4}}{(4\chi v_0(\chi)+\chi-1)^{4}}+\\
\left.\frac{\left.8\chi^{2} v_0(\chi)^{3}+12\chi^{2} v_0(\chi)^{2}-4\chi^{2} v_0(\chi)-6\chi v_0(\chi)^{2}+2 v_0(\chi)^{2}+6\chi v_0(\chi)-2 v_0(\chi)\right)}{(4\chi v_0(\chi)+\chi-1)^{4}}\,d\chi\right)
\end{aligned}\label{v1sol}
\end{equation}
where $C$ is an integration constant and
\begin{equation}
\begin{aligned}K\left(\bar{\varrho}\right)=\int_{1}^{\bar{\varrho}}-2\left(\frac{64\chi^{3} v_0(\chi)^{4}+64\chi^{3} v_0(\chi)^{3}+4\chi^{3} v_0(\chi)^{2}-6\chi^{3} v_0(\chi)-64\chi^{2} v_0(\chi)^{3}-24\chi^{2} v_0(\chi)^{2}}{(4\chi v_0(\chi)+\chi-1)^{4}}+\right.\\
\left.\frac{10\chi^{2} v_0(\chi)+20\chi v_0(\chi)^{2}-2\chi v_0(\chi)-2 v_0(\chi)-\chi^{3}+3\chi^{2}-3\chi+1}{(4\chi v_0(\chi)+\chi-1)^{4}}\,d\chi\right).
\end{aligned}\label{K}
\end{equation}
Replacing $v_0$ in Eq. (\ref{v1sol})  by its Taylor expansion  (\ref{v0}) yields:

\begin{equation}
\begin{aligned}
v_1\left(\bar{\varrho}\right)=-\tau -h \tau  (\alpha -20 \tau +4)+h^2 \left(\tau ^2 (8 \alpha -156 \tau +37)+\tau  (\alpha +12 \tau  (8 \tau -3)) \log (h)+C \right)+O\left(h^3\log(h)\right).
\end{aligned}\label{log-term}
\end{equation}
The analyticity of $v_1$ implies that the  log term in Eq. (\ref{log-term}) is absent. This requires that its prefactor vanishes, that is, $\alpha= 36 \tau -96 \tau^2$ which is the same as Eq. (9) of the main text. To all orders checked (up
to 5th order) this also eliminates the following log terms. 

Notice that the expression (\ref{v1sol}) giving $v_1(\bar\varrho)$ is ill-conditioned for a numerical plot of this function because of the poles of the integrands of  $K$ in Eq.~(\ref{K}) and in $v_1$ in Eq.~(\ref{v1sol}). Although the final expression for $v_1$ is well-defined it is tricky to get rid of apparent divergencies showing up because of the poles within the integrands: This requires adding and subtracting divergencies and making some integration by parts. For this reason, it is simpler to numerically integrate Eq.~(\ref{v1dif}).

\section{Singular perturbation theory and computation of the boundary layers of the singular FPs $SA_3$ and $S\tilde{A}_4$}

We now show how to compute the boundary layers of the singular FPs $SA_3$ and $S\tilde{A}_4$.
We  use the following change of variables: $\ensuremath{\bar{\varrho}=\frac{\tilde{\varrho}}{N}+\bar{\varrho}_{0}}$,
$\ensuremath{\bar{V}=\frac{\tilde{V}}{N}+\bar{V}_{0}}$, where the point $(\bar{\varrho}_{0},\bar{V}_{0})$ corresponds to the location of the cusp on Fig.(2) of the main text. As we are working here to the lowest order in $\epsilon$, one may check that $d(N)=3-\alpha/N$ can be replaced in Eq.(\ref{eq:V})  by $d(N)=3$. Moreover, the potential is linear: $\bar{V}=\bar{\varrho}$ on
the left of the cusp as in Fig.(2) of the main text, thus we also have at the position of the cusp: $\bar{V}_{0}=\bar{\varrho}_{0}$. Inserting all of these elements in Eq.(\ref{eq:V}), we obtain to leading order in $\epsilon$:

\begin{equation}
0=1-3\bar{\varrho}_{0}+\bar{\varrho}_{0}\,\tilde{V}'+2\,\bar{\varrho}_{0}\,\tilde{V}'^{^{2}}-\,\tilde{V}'-2\,\bar{\varrho}_{0}\,\tilde{V}''
\end{equation}
whose solution is 
\begin{equation}
\tilde{V}'\left(\tilde{\varrho}\right)=\frac{(5\text{\ensuremath{\bar{\varrho}_{0}}}-1)\tanh\left(\frac{(1-5\text{\ensuremath{\bar{\varrho}_{0}}})\left(\tilde{\varrho}-C\right)}{4\text{\text{\ensuremath{\bar{\varrho}_{0}}}}}\right)-\text{\ensuremath{\bar{\varrho}_{0}}}+1}{4\text{\text{\ensuremath{\bar{\varrho}_{0}}}}}
\end{equation}
where $C$ is some integration constant.

Moreover, using (\ref{eq:V}) for $1/N=\epsilon=0$ and $d=3$ without the previous change of variables $\ensuremath{\bar{\varrho}=\frac{\tilde{\varrho}}{N}+\bar{\varrho}_{0}}$,
$\ensuremath{\bar{V}=\frac{\tilde{V}}{N}+\bar{V}_{0}}$, 
we obtain: 
\begin{equation}
\bar{V}'(\bar{\varrho}) \left(\bar{\varrho}+2 \bar{\varrho}\bar{V}'(\bar{\varrho})-1\right)-3 \bar{V}(\bar{\varrho})+1=0.\label{polyV}
\end{equation}
By evaluating Eq. (\ref{polyV}) at $\bar\varrho=\bar\varrho_0=\bar V_{0}$, we obtain two possible values for $\bar{V}'(\varrho_{0})$:
\begin{equation}
\begin{cases}
\bar{V}_{-}'(\bar{\varrho}_{0})=1\\
\bar{V}_{+}'(\bar{\varrho}_{0})=\frac{1}{2}\left(\frac{1}{\bar{\varrho}_{0}}-3\right).
\end{cases}
\end{equation}
These are the two distinct derivatives at the cusp with $\bar{V}_{-}'(\bar{\varrho}_{0})=\bar{V}'(\bar{\varrho}_{0}^{-})$
and $\bar{V}_{+}'(\bar{\varrho}_{0})=\bar{V}'(\bar{\varrho}_{0}^{+})$.
Using this we may rewrite $\tilde{V'}\left(\tilde{\varrho}\right)$ as
\begin{equation}
\tilde{V'}\left(\tilde{\varrho}\right)=\frac{-(\bar{V}'(\bar{\varrho}_{0}^{-})-\bar{V}'(\bar{\varrho}_{0}^{+}))\tanh\left(\frac{(\bar{V}'(\bar{\varrho}_{0}^{-})-\bar{V}'(\bar{\varrho}_{0}^{+}))\left(\tilde{\varrho}-C\right)}{2}\right)+\bar{V}'(\bar{\varrho}_{0}^{+})+\bar{V}'(\bar{\varrho}_{0}^{-})}{2}.
\end{equation}

Finally we choose $C=0$ such that for $\bar{\varrho}=\bar{\varrho}_{0}$ (and thus $\tilde{\varrho}=0$)
we have $\tilde{V'}\left(0\right)=\frac{\bar{V}'(\bar{\varrho}_{0}^{+})+\bar{V}'(\bar{\varrho}_{0}^{-})}{2}$

\end{document}